\documentclass[journal=acsodf,manuscript=article]{achemso}

\usepackage{amsmath,amssymb}
\usepackage{graphicx}
\usepackage{tabularx}
\usepackage{multirow}
\usepackage{makecell}
\usepackage{xcolor}
\usepackage{placeins}

\providecommand{\colrule}{\hline}
\newsavebox{\ruledbox}
\newenvironment{ruledtabular}%
  {\begin{lrbox}{\ruledbox}}%
  {\end{lrbox}\resizebox{\ifdim\wd\ruledbox>\linewidth \linewidth\else \wd\ruledbox\fi}{!}{\usebox{\ruledbox}}}

\makeatletter
\AtBeginDocument{%
  \let\acs@orig@maketitle\maketitle
  \renewcommand\maketitle{\begingroup\singlespacing\acs@orig@maketitle\endgroup}%
}
\makeatother

\newif\ifhighlight
\highlightfalse
\newcommand{\rev}[1]{\ifhighlight\textcolor{blue}{#1}\else#1\fi}

\author{Nikoletta Jegenyes}
\affiliation{HUN-REN Wigner Research Centre for Physics, P.O.\ Box 49, H-1525 Budapest, Hungary}
\author{Vladimir Verkhovlyuk}
\affiliation{HUN-REN Wigner Research Centre for Physics, P.O.\ Box 49, H-1525 Budapest, Hungary}
\author{Szabolcs Czene}
\affiliation{HUN-REN Wigner Research Centre for Physics, P.O.\ Box 49, H-1525 Budapest, Hungary}
\alsoaffiliation{Doctoral School on Materials Sciences and Technologies, \'Obuda University, B\'ecsi \'ut 96/b, H-1034 Budapest, Hungary}
\author{Attila Cs\'aki}
\affiliation{HUN-REN Wigner Research Centre for Physics, P.O.\ Box 49, H-1525 Budapest, Hungary}
\author{S\'andor Kollarics}
\affiliation{HUN-REN Wigner Research Centre for Physics, P.O.\ Box 49, H-1525 Budapest, Hungary}
\alsoaffiliation{SOLEIL Synchrotron, L'Orme des Merisiers, RD128, Saint Aubin, France, 91190}
\alsoaffiliation{Department of Physics, Institute of Physics, Budapest University of Technology and Economics, M\H{u}egyetem rakpart 3., H-1111 Budapest, Hungary}
\author{Olga Krafcsik}
\affiliation{Department of Atomic Physics, Institute of Physics, Budapest University of Technology and Economics, M\H{u}egyetem rakpart 3., H-1111 Budapest, Hungary}
\alsoaffiliation{Institute of Technical Physics and Materials Science, HUN-REN Centre for Energy Research, P.O.\ Box 49, H-1525 Budapest, Hungary}
\author{Zsolt Czig\'any}
\affiliation{Institute of Technical Physics and Materials Science, HUN-REN Centre for Energy Research, P.O.\ Box 49, H-1525 Budapest, Hungary}
\author{David Beke}
\email{beke.david@wigner.hun-ren.hu}
\affiliation{HUN-REN Wigner Research Centre for Physics, P.O.\ Box 49, H-1525 Budapest, Hungary}
\alsoaffiliation{Kand\'o K\'alm\'an Faculty of Electrical Engineering, \'Obuda University, Tavaszmez\H{o} u. 17., H-1084 Budapest, Hungary}
\author{Adam Gali}
\email{gali.adam@wigner.hun-ren.hu}
\affiliation{HUN-REN Wigner Research Centre for Physics, P.O.\ Box 49, H-1525 Budapest, Hungary}
\alsoaffiliation{Department of Atomic Physics, Institute of Physics, Budapest University of Technology and Economics, M\H{u}egyetem rakpart 3., H-1111 Budapest, Hungary}
\alsoaffiliation{MTA--WFK Lend\"{u}let "Momentum" Semiconductor Nanostructures Research Group, P.O.\ Box 49, H-1525 Budapest, Hungary}

\title{Materials and spin characteristics of nanodiamonds partially covered with amino groups and embedded with nitrogen-vacancy color centers}
\abbreviations{FND,NV,ODMR,ESR,FTIR,XPS,ZFS}

\begin{document}

\begin{abstract}
Fluorescent nanodiamonds (FNDs) with optically read qubits hold great potential for detecting electric and magnetic fields, temperature, and other nanoscale physicochemical quantities relevant to chemistry and biology. Proper surface functionalization is essential for their application as probes, but surface modifications can impact qubit sensor properties. In this work, we systematically study nitrogen-vacancy (NV) color centers in FNDs as a function of size and surface termination. FNDs were produced from high-pressure, high-temperature diamonds, with NV centers introduced via electron irradiation and annealing. The initial oxygen-covered FNDs were homogenized with hydroxyl ($-$OH) groups as reference samples, while the noninvasive Hofmann degradation introduced amino ($-$NH$_2$) groups for potential direct biomolecule attachment. Although amino groups may not homogeneously cover these nanodiamonds, we simply label them as $-$NH$_2$ terminated FNDs in this context. We monitored charge state stability and the zero-field splitting parameters of the embedded NV centers. This study yields two principal advances. First, we resolve the size dependence of the NV($-$) zero-field splitting parameters across the 10$-$140~nm range and show that the symmetry-breaking $E$ parameter decreases monotonically from $\sim8$ to $\sim5$~MHz with increasing size while the axial $D$ parameter is shifted only in the smallest ($\leq30$~nm) particles, thereby disentangling the static-strain and fluctuating electric-field contributions to the spin levels. Second, while NV charge state stabilization was observed in both $-$OH- and $-$NH$_2$-terminated FNDs above a certain size, we demonstrate that a remarkably high and laser-power-independent NV($-$) content ($f_{\text{NV}(-)}\approx0.8$) is achieved by wet-chemical Hofmann amino termination only in 140~nm particles, an effect we link through electron spin resonance to the degradation of surface paramagnetic defects rather than to the introduction of new ones.
\end{abstract}

\section{Introduction}
\label{sec:introduction}

Nitrogen-vacancy (NV) centers in nanodiamonds are widely explored as nanoscale quantum sensors for chemical and biological applications~\cite{Kucsko2013, Rendler2017, Fujisaku2019, Webb2020, Liu2023}. Their electronic spin can be optically polarized and read out via optically detected magnetic resonance (ODMR), enabling magnetic, electric, and thermal sensing at ambient conditions~\cite{Doherty2013, Barry2020}. For biological use, NV centers must be incorporated into nanodiamonds (fluorescent nanodiamonds, FNDs) whose surfaces can be chemically functionalized without degrading spin performance.

In nanodiamonds, NV centers typically reside close to the surface. As a result, their charge stability and spin relaxation times are highly sensitive to surface chemistry~\cite{Tisler2009, Rondin2010, Ryan2018}. Surface-induced band bending, charge traps, and paramagnetic defects may promote charge switching between NV($-$) and NV($0$), modify fluorescence stability, and shorten spin relaxation times~\cite{Pfender2017, Broadway2018, Freire-Moschovitis2023}. Thus, controlling the surface termination is essential for preserving both the optical and quantum properties of NV centers in FNDs.

Recent studies have shown that tailored surface chemistry can significantly influence spin coherence and stability. Oxygen-, hydrogen-, and mixed nitrogen/oxygen terminations have been investigated as routes to mitigate surface-related noise~\cite{Kaviani2014, Sangtawesin2019, Malkinson2024, abendroth2022}. In particular, amino-functionalized surfaces are attractive for biological applications due to their straightforward bioconjugation chemistry~\cite{Coffinier2007}. However, protonated amino groups may induce negative electron affinity and affect the charge stability of near-surface NV centers~\cite{Zhu2016, Kawai2019}. Therefore, understanding how amino termination influences NV optical and spin properties is critical.
\begin{figure} 
\includegraphics[width=0.5\textwidth]{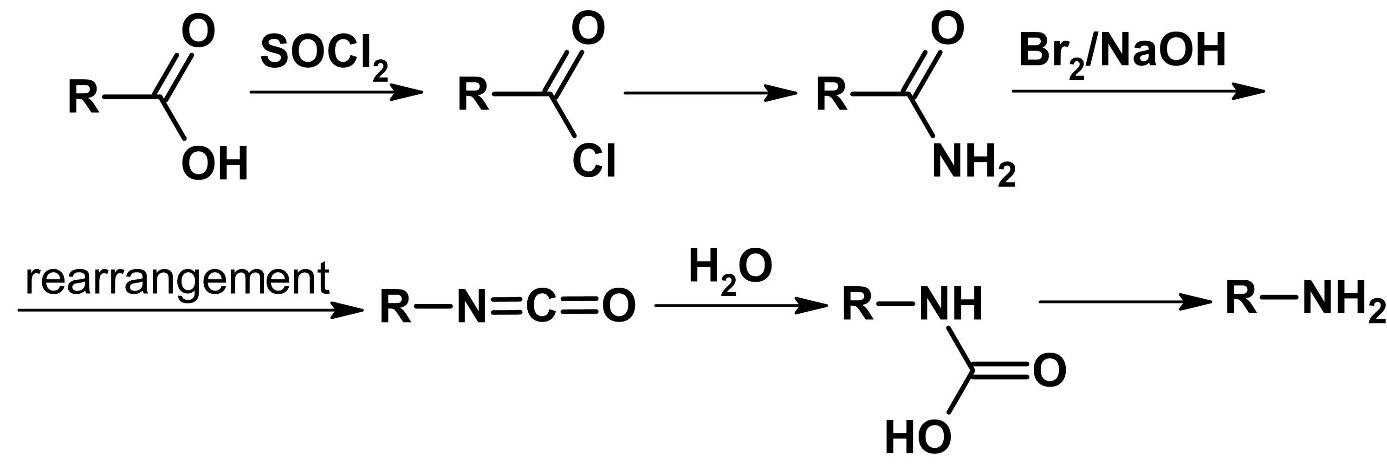}
    \caption{\label{fig:Hofmann}Hofmann degradation mechanism.}
\end{figure}
Amino termination of diamond surfaces is commonly achieved using plasma-based methods~\cite{koch2011, laube2017, abendroth2022}, which require specialized equipment and careful process control. Here, we instead employ the Hofmann degradation mechanism~\cite{Hofmann-mechanism} (Figure~\ref{fig:Hofmann}) as a wet-chemical route to introduce amino groups onto nanodiamond surfaces. This approach selectively modifies oxygen-containing surface functionalities while avoiding graphitization and particle aggregation~\cite{Dai2016, Czene2023}. The method has previously been successfully applied to other oxygen-terminated nanomaterials~\cite{Czene2023}.

The individual phenomena probed here have been examined separately in earlier work. The NV($-$) charge-state fraction in FNDs is known to grow with particle size for oxygen-terminated single centers~\cite{Rondin2010} and ensembles~\cite{Wilson2019}; the size of milled nanodiamonds governs the relative weight of the \rev{substitutional-nitrogen donor (P1) center}~\cite{Loubser1978} and the milling-induced $S=1/2$ surface signal in the $g\approx2$ electron spin resonance (ESR) spectrum~\cite{Shenderova2019, shames2015, panich2015}, and wet-chemical or plasma surface treatments are known to modify the spin properties of near-surface NV centers~\cite{Ryan2018, Kaviani2014, abendroth2022, Malkinson2024, Gulka2024}. What is still lacking is a unified picture that ties these effects to the zero-field splitting (ZFS) parameters of the embedded NV($-$) ensemble across a single, wide size series prepared with identical surface chemistry, together with a noninvasive amino-termination route that preserves, rather than degrades, the NV charge state.

In this work, we address these questions by systematically investigating commercially available HPHT-derived FNDs (10--140~nm) with oxygenated, reduced ($-$OH-terminated), and amino-terminated surfaces, characterized on the same sample series by FTIR, X-ray photoelectron spectroscopy (XPS), zeta-potential, photoluminescence (PL), ESR, and ODMR measurements. We report two main results. First, we resolve the NV($-$) ZFS $D$ and $E$ parameters as a continuous function of FND size and find that, while $E$ decreases monotonically with size for every termination, $D$ departs from its bulk value only in the smallest ($\leq30$~nm) particles; following the analysis of ref~\citenum{Mittiga2018} and the recent \textit{ab initio} result of ref~\citenum{Pershin2025}, this separates a static-strain contribution that dominates below $\sim30$~nm from a fluctuating electric-field contribution of parasitic defects that scales with the NV-to-surface distance. Second, we show that the Hofmann degradation provides a noninvasive, wet-chemical amino termination that suppresses surface paramagnetic defects rather than creating them, as corroborated by photoESR, yielding a high and laser-power-independent NV($-$) fraction in 140~nm particles. Taken together, and given their mutual consistency in the ESR analysis, these results indicate that Hofmann-based amino termination can provide a surface suitable for biological functionalization while preserving---and, in the largest particles, enhancing---the NV charge-state and spin characteristics.

\section{Methods}
\label{sec:methods}
FNDs purchased from Ad\'amas Nanotechnologies Inc.\ were used to study the effects of surface terminations. These FNDs were derived from high-pressure, high-temperature (HPHT) diamonds, which were milled to produce nanodiamonds~\cite{Shenderova2019}. The nanodiamonds were then electron-irradiated and annealed to generate NV defects. Subsequently, the nanodiamond surfaces were cleaned and oxygenated~\cite{Shenderova2019}.

Our analysis (see below) revealed that the FND solutions contain sulfur, which may originate from their preparation method. 
The FNDs were either used as is or washed before measurements by applying solvent exchange centrifugation with \rev{deionized (DI) water} three times, acetone twice, and DI water five times, to remove sulfur-related residues that affected the optical properties of the samples. 

Hydroxyl-terminated FNDs were synthesized via LiAlH$_4$ reduction, adapted from ref~\citenum{Shenderova2011}. Briefly, the solvent in the aqueous FND solution was replaced with anhydrous tetrahydrofuran (THF), dried over CaCl$_2$, and removed by centrifugation. Subsequently, 1.6~mL of a 1.0~mol/L LiAlH$_4$ solution was added, resulting in a final LiAlH$_4$ concentration of 0.2~mol/L. The reaction mixture was refluxed overnight under vigorous stirring. Following THF evaporation, the reaction was quenched by adding 1~mol/L HCl. Purification and collection were performed via centrifugation using tubes with poly(ether sulfone) filter membranes. The product was washed multiple times with DI water until a neutral pH was reached, followed by sequential washes with acetone and DI water to ensure complete removal of residual contaminants. It should be noted that the applied method does not produce a uniformly OH-terminated surface. Rather, the strong reducing agent employed increases the concentration of $-$OH moieties while simultaneously reducing the concentration of functional groups in higher oxidation states.

Amino-terminated FNDs were prepared following a modified procedure based on ref~\citenum{Czene2023}. Initially, the FND suspension was dried, and SOCl$_2$ was added, followed by sonication at 50~$^\circ$C for 3~h. The SOCl$_2$ was then removed under vacuum at 50~$^\circ$C, after which an NH$_3$/dioxane solution was introduced. The mixture was sonicated again at 50~$^\circ$C, then heated to 80~$^\circ$C for 10~min in the presence of NaOH and Br$_2$. To eliminate excess Br$_2$, NH$_3$ was added to the cooled solution, and any residual NH$_3$ was removed via vacuum evaporation at 50~$^\circ$C. The pH was adjusted using HCl, and purification was carried out through ten cycles of centrifugation and redispersion in water. It should be noted that the applied reactions primarily target $-$COOH groups and related oxygen-containing surface moieties (e.g., hydroxyl groups), converting them first to $-$Cl and subsequently to $-$NH$_2$. Nevertheless, the complete substitution of all O-related functional groups with nitrogen-containing ones is not expected to occur: as a result, they both coexist on the surface.

We called the initial samples as-received, whereas they were labeled as "washed" when the sulfur-related residues were removed. We labeled the particles with different types of surface termination as follows: for hydroxyl-termination, "NDx$-$OH" and for amino-termination,"NDx$-$NH$_2$" were used, where "x" corresponds to the size of the FND. The considered samples are listed in Table~\ref{tab:FND}.

    \begin{table}[h!]
        \caption{\label{tab:FND}Data of the applied FNDs with giving the sample label, the average diameter of the particle ($d$), the density of the NV centers ($\rho$), the number of NV centers per particle (\#NV), and the density of the P1 centers ($\rho_{\text{P}_1}$). Data are provided by Ad\'amas Nanotechnologies Co. (Raleigh, NC, USA) and presented in Ref~\citenum{Shenderova2019}.} 
        \begin{ruledtabular}
            \begin{tabular}{c|cccc}
                Label  &  $d$ (nm) & $\rho$ (ppm) & \#NV & $\rho_{\text{P}_1}$ (ppm) \\
                \hline
                ND140 & 140  & $\leq$ 3 & 385 & $\sim$ 8  \\
                ND90  & 90 & $\sim$ 3 & 200 & -    \\
                ND70  & 70 & $\leq$ 3 & 95 & $\sim$ 4.8    \\
                ND50  & 50  & $\leq$ 3 & 23& $\sim$ 4   \\
                ND40  & 40  & $\leq$ 2  & 12-14 &   $\sim$ 2    \\
                ND30  & 30  & $\leq$ 2 & 5-6 & -    \\
                ND10  & 10  & $<$ 1  & 1-2 & -    \\
            \end{tabular}
        \end{ruledtabular}
    \end{table}

\rev{Scanning electron microscopy (SEM; TESCAN MIRA3) and high-resolution transmission electron microscopy (HRTEM; Thermo Fisher Themis C$_s$-corrected transmission electron microscope, TEM)} were used to observe the shape and size of FNDs. \rev{SEM combined with energy-dispersive X-ray spectroscopy (SEM-EDS; spectra recorded by an Element EDS system)} and HRTEM  methods were used for basic element analysis.  In the case of EDS measurements, the particles were dispersed over a Si substrate. For the HRTEM analysis, the aqueous suspension of FNDs was drop-cast onto a TEM grid (TED Pella) covered with an ultrathin carbon layer supported by lacey carbon after the purification.

Two methods were used to characterize the success of the surface modification. The samples were analyzed using Fourier transform infrared spectroscopy (FTIR) (we applied either a Bruker IFS66 spectrometer or a Bruker Tensor 37 spectrometer equipped with \rev{deuterated triglycine sulfate (DTGS)} detectors) where the vibration bands of the surface groups are monitored. The concentration of the colloidal diamond samples was adjusted to 0.5\,mg/mL, and 50~$\mu$L was drop-cast onto a silicon wafer to ensure uniform sample thickness and quantity across all samples.

We further analyzed the chemical bonds of selected as-received and amino-terminated FNDs by surface-sensitive X-ray photoelectron spectroscopy (XPS). The XPS measurements were carried out using a twin anode X-ray source (Thermo Fisher Scientific, Waltham, MA, USA, XR4) and a hemispherical energy analyzer with a nine-channel multichanneltron detector (SPECSGROUP, Berlin, Germany, Phoibos 150 MCD). The base pressure of the analysis chamber was around $2 \times 10^{-9}$~mbar. Samples were analyzed using an Mg K$\alpha$ (1253.6~eV) anode without monochromatization. The FND samples were 
drop-cast onto a niobium substrate.

Zeta potential measurements were carried out with Malvern instrument (Malvern Zetasizer Nano Z, Cambridge, United Kingdom) in an aqueous medium (pH 6) at 25~$^\circ$C in all cases. The zeta potential
values were recorded from 3 measurements, and each measurement consisted of 45 runs. The samples were in a 10~mm polystyrene cuvette equipped with palladium electrodes.

The photoluminescence (PL) spectra were detected using a Renishaw inVia Raman Microscope with a 50$\times$ Leica objective. The PL spectra were measured on drop-cast FNDs on a Si surface. The excitation wavelength was 532~nm and the power of the laser source was adjusted within 1-800~$\mu$W range and calibrated under the microscope objective. The power range corresponds to 125-2654~$\mu$W/cm$^{2}$ irradiance. The largest applied power is an upper bound limit in our measurements, as a larger power of the laser results in such a high intensity of emission from our FND samples that could be detrimental for the operation of the detectors. For the accuracy of the analysis, we have repeated each measurement 5-7 times on all samples by using the same excitation power. Several measurements were performed using different laser powers on the same area of the samples.

We studied X-band electron spin resonance (ESR) on a Magnettech MiniScope MS-400 X-band spectrometer equipped with a TE$_102$ rectangular resonator. The samples were placed in high quality, defect free quartz tubes. Care was taken to employ a low microwave (MW) exciting power (0.2~mW) and a low magnetic field (0.15-0.25~G) modulation to avoid any distortion to the ESR line shapes. The measurements were carried out by applying 520~nm laser illumination at room temperature to mimic the conditions of the optically detected magnetic resonance (ODMR) studies. 
The Matlab-based software EasySpin~\cite{Stoll2006} was used to simulate the measured ESR data. The P1 center with known spin Hamiltonian parameters was used for calibration of the MW frequency. In the simulation, the $g$ value of the P1 center was fixed, and an anisotropic strain was applied to the hyperfine interaction in order to reproduce the P1 line shape in the spectra. In the initial fitting process, Voigt broadening was tested for the central line; however, Lorentzian broadening was ultimately employed because it provided the best description of the observed peaks. The $g$-factors for the central line were allowed to vary during the fitting procedure.

Continuous wave (cw) ODMR 
measurements were conducted at constant excitation power. The zero-field splitting (ZFS) $D$ and $E$ parameters of the observed ensemble NV centers were monitored by cw-ODMR measurements at room temperature (297~K) where $D$ separates the $m_S=0$ and the average of the $m_S=-1$ and $m_S=+1$ levels, $2\times E$ separates the $m_S=-1$ and $m_S=+1$ levels. In other words, $D-E$ and $D+E$ dips appear in the cw-ODMR spectrum at zero magnetic field. We note that $D=2.87$~GHz and $E=0$ in the case of a perfect diamond, while nonzero $E$ occurs as a result of strain or fluctuating electric fields.
A 520~nm laser served as the excitation source, delivering 10~mW power before the \rev{0.9 numerical aperture (NA)} Carl Zeiss objective. Emission was collected through the same objective. The collected emission was then passed through a 550~nm dichroic mirror (Semrock) and a 650~nm long-pass filter (Thorlabs) to an \rev{avalanche photodiode (APD)} detector (Excelitas SPCM-AQRH-44). The output of the APD detector was connected to the input of a lock-in amplifier (SR830).

For the microwave field, a Vaunix LabBrick LSG-402 source, a high-power amplifier (Mini-Circuits ZHL-16W-43+), a Mini-Circuits ZASWA-2-50DRA+ switch, and a microwave antenna under the sample were employed. Microwave power ranged from 20 to 40~dBm. The FND colloidal solutions were drop-cast onto a nonfluorescent borosilicate cover glass of type No.0 ($\approx0.10$~mm thickness) and positioned close to the antenna. The aforementioned optical system creates a $\sim300$~nm sampling area, therefore, each measurement was done on multiple NDs and on NV center ensembles. We have repeated the analysis at least three times on different regions of the drop-cast samples. Evaluation of cw-ODMR results was performed using EasySpin simulation software in a powder spectrum mode (i.e., assuming that the spin center orientations are randomly but uniformly distributed in the sample)~\cite{Stoll2006}.

\rev{A continuous-wave rather than a pulsed read-out was chosen deliberately for this class of samples. The quantities reported in this work, the ZFS parameters $D$ and $E$, are static spin-Hamiltonian parameters that are read off the \textit{positions} of the resonances; they are properties of the ground-state spin levels and are therefore the same whether the resonance is recorded in continuous-wave or in pulsed mode. A pulsed scheme would in addition require a well-defined microwave rotation angle, which cannot be realized in a drop-cast nanodiamond powder: the particles are randomly oriented, so the Rabi frequency of each NV center depends on the projection of the driving field onto its own $C_{3v}$ symmetry axis; it varies further with the inhomogeneity of the microwave field over the illuminated area, and with the strain- and electric-field-dependent transition matrix elements that also give rise to the spread of $E$ reported below. The resulting broad distribution of Rabi frequencies renders a nominal $\pi$ pulse ill-defined across the ensemble and suppresses the coherent contrast, whereas cw-ODMR is insensitive to pulse calibration and yields directly the orientation-averaged powder spectrum that we analyze with EasySpin~\cite{Stoll2006, Barry2020}.}

\section{Results and Discussion}
\label{sec:results}

We first report the basic morphological and structural characterization of the FND samples, and then we continue the analysis of the surface termination of FNDs upon various treatments. 

SEM images reveal that the FND particles exhibit an irregular shape with distinct crystal facets in all sizes (see Figure~S1), which are consistent with prior studies~\cite{Shenderova2019}. 
The size of these nanodiamonds is usually measured as the largest distance between the facets of the particle. Due to the nonspherical characteristics of the particles,  the distance of the NV defects from the nanodiamond surface cannot be well estimated, and it is most likely much shorter than the half-size of the particle. Furthermore, the average distance of NV defects from the surface with growing size of nanodiamonds could show a monotonous function, but with a relatively gentle slope.

The success of surface modification was evaluated using FTIR and XPS fine structure analysis, while elemental composition was investigated by XPS, SEM-EDS, and HRTEM-EDS on selected FND samples. 
Zeta potential measurement will provide additional evidence of surface modification and colloid stability.

    \begin{figure}[h]
        \includegraphics[width=0.5\textwidth]{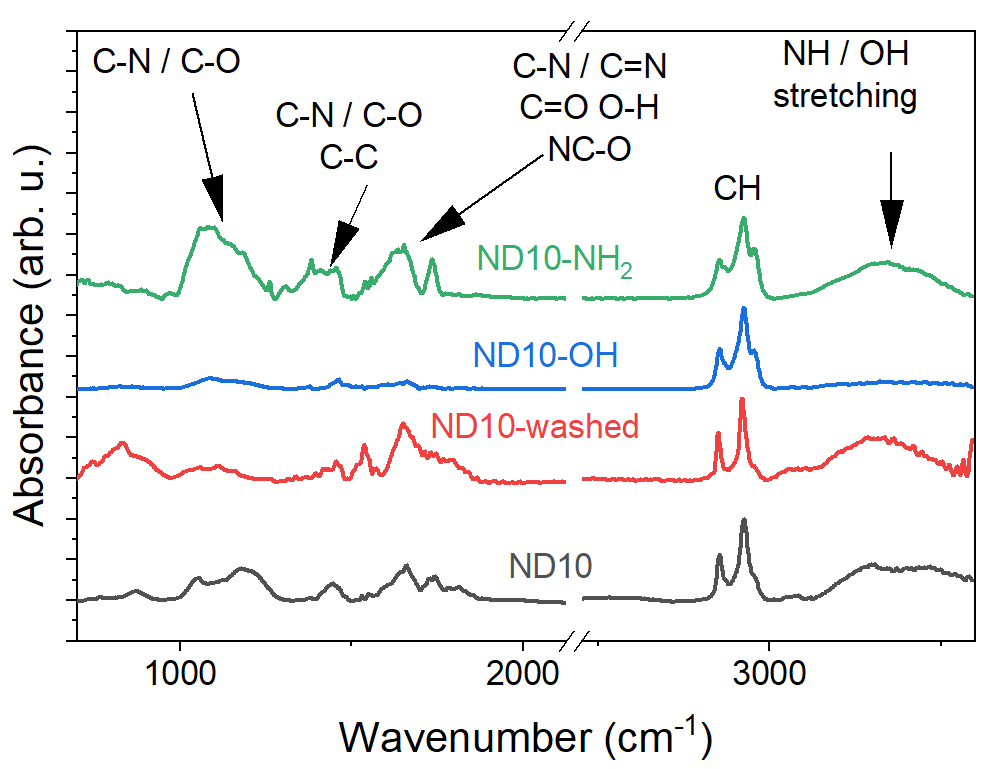}
        \caption{\label{fig:IR-ND10}Infrared vibration spectra of the as-received, washed, $-$OH and $-$NH$_{2}$ terminated ND10 FNDs.}
    \end{figure}

To enable a meaningful comparison among the FTIR spectra of the as-received, cleaned, and surface-modified samples, all spectra were normalized to a [0.1] intensity. This normalization was necessary because no FTIR peak remains unchanged throughout the reactions.
    
The FTIR spectra of the as-received samples indicated the presence of oxygen-containing groups that we show on the exemplary ND10 samples in Figure~\ref{fig:IR-ND10}. We note that very similar features were observed for the other FND sizes (Figure~S4) where we list the detailed analysis in Tables~S1--S7, with the band assignments based on refs~\citenum{Coates2006, Ryan2018, Petit2018}.

We identified oxygen-related functional groups, including the C$-$O stretching mode at 1062~\(\text{cm}^{-1}\) and a broad band above 3130~\(\text{cm}^{-1}\) in the as-received FNDs, originate from O$-$H stretching in $-$COOH and $-$OH groups.   
The C=O stretching vibrations of carboxyl ($-$COOH) groups appear at 1773 and 1660~\(\text{cm}^{-1}\), providing clear evidence of carboxyl functionalities on the oxygenated diamond surface. These carboxyl groups likely play a dominant role in inducing band bending~\cite{Freire-Moschovitis2023}. The weak peak at 1332~\(\text{cm}^{-1}\) likely corresponds to the C$-$C stretching vibration of the diamond lattice.


Comparing the FTIR spectra of as-received and washed FNDs, two major changes are affirmed:  
(i) the C$-$H bending modes \cite{Petit2018} at 1460~\(\text{cm}^{-1}\)~ decrease in intensity after washing, and  
(ii) the peaks around 1216~\(\text{cm}^{-1}\) show a reduction.  
The last-mentioned spectral region may correspond to both C$-$O stretching vibrations and sulfonate groups. The SEM-EDS and HRTEM-EDS analysis have revealed that, besides the presence of carbon, oxygen, and Si (originated from the substrate), the as-received samples also contain sulfur (see Figures~S2 and S3). Since the latter observation was made only for the as-received samples, but not for the cleaned ones, we attribute this band primarily to sulfonate groups. 
The removal of these sulfur-containing residues hindered redispersion in water and also led to changes in the optical properties of the nanodiamonds. As the surface modification steps used in this study inherently include solvent exchange and washing processes, the cleaned samples were considered more suitable as a reference for evaluating the optical properties reported herein. 

The intensity of the C=O stretching vibrations at $1773$ and $1660$~\(\text{cm}^{-1}\) is reduced and the O$-$H stretching band is narrowed in ND-OH samples (blue curve in Figure~\ref{fig:IR-ND10}), as a result of chemical reduction of the surface groups. It is apparent that the C=O stretching vibration modes did not fully vanish after reduction which implies that the FND surface is not homogeneously terminated with $-$OH groups. Achieving homogeneity is challenging due to the diverse surface moieties requiring different chemistry for conversion, the unknown effect of facets on the chemical reactions, steric hindrance due to the irregular particle shape, and other factors. Nevertheless, these results are in good agreement with a previous study~\cite{Shenderova2011}. 

For ND-NH$_2$ samples, FTIR spectra showed the reduction of all the oxygen-related peaks and the emergence of nitrogen-related ones (c.f. green and red curves in Figure~\ref{fig:IR-ND10}). Analysis of the FTIR spectra showed the presence of the torsional vibration of the amino-group at $750-800$~cm\(^{-1}\) as well as the C-N stretching mode of primary amines, which corresponds to the peaks in the $1060-1090$~cm\(^{-1}\) region. The secondary amino-vibration can be found between 1130 and 1190~cm\(^{-1}\), and the signal of the tertiary amino-groups is between 1160 and 1210~cm\(^{-1}\). Furthermore, N-H bending vibration band at around 1540-1650~cm\(^{-1}\) also occurs in the FTIR spectra of FNDs after Hofmann degradation (see also Table~S1). This is a good indication that the amino-termination of FNDs was indeed achieved. The same principle applies for amino-terminated FNDs as to the hydroxyl-terminated FNDs, i.e., the surface termination is incomplete, and oxygen-related peaks remain at the surface of FNDs. The presence of primary and secondary amines also indicates inhomogeneity, even for the nitrogen-related groups.

\begin{figure}[h]
        \includegraphics[width=0.5\textwidth]{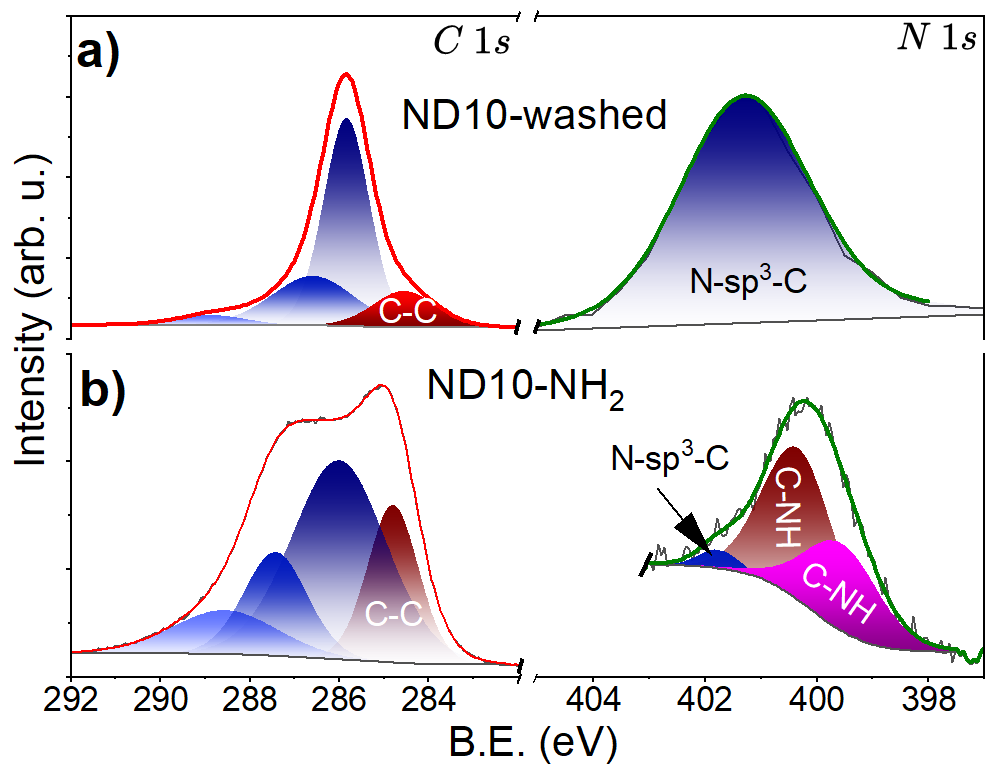}
        \caption{\label{fig:XPSa} Normalized C~1\textit{s} and N~1\textit{s} fine structure XPS spectra of (a) washed (b) and $-$NH$_2$ terminated FNDs. The surface termination-related binding energies are colored blue on the C~1\textit{s} spectra.}
    \end{figure}

To supplement the FTIR analysis, the C~1\textit{s} and N~1\textit{s} fine structure spectra were investigated by XPS. It has shown distinct differences between the washed and amino-terminated samples (see Figure~\ref{fig:XPSa} and Figure~S5). The C~1\textit{s} spectra exhibited a notable increase in the C--C bond-related peak intensity relative to other surface termination-related components, consistent with previous reports \cite{motlag2020, singh2016, auciello2023, rodgers2024, arnault2018}. However, the identification of C--N bonds in the C~1\textit{s} region remains ambiguous due to the spectral overlap with C--O bond binding energies. The N~1\textit{s} spectra revealed a peak at approximately 404~eV in some samples, attributed to nitrite residues~\cite{auciello2023}. 
Furthermore, a consistent peak at around 401~eV was observed in all samples, corresponding to N--sp$^3$--C bonds associated with nitrogen vacancies in the nanodiamond lattice~\cite{auciello2023, osipov2018, osipov2019, motlag2020}. Upon amino functionalization, a new peak emerged near 400~eV, which is attributed to the formation of C--N--H bonds~\cite{auciello2023}, indicating successful surface modification (Table~\ref{tab:XPS}).



    \begin{table}[h]
        \caption{\label{tab:XPS}XPS results on ND10, ND50, and ND90 FNDs showing the atomic concentrations of C, O, and N, and the N/C and N/O atomic ratios.}
        \begin{ruledtabular}
            \begin{tabular}{l|ccccc}
                Sample  & C (at\%) & O (at\%) & N (at\%) & N/C & N/O \\
                \hline 
                ND10-ref        & 85.43 & 13.48 & 1.08 & 0.013 & 0.080 \\
                ND10-NH$_{2}$   & 87.32 & 10.93 & 1.75 & 0.020 & 0.160 \\
                ND50-ref        & 89.12 & 10.29 & 0.59 & 0.007 & 0.057 \\
                ND50-NH$_{2}$   & 93.70 &  4.00 & 2.30 & 0.025 & 0.575 \\
                ND90-ref        & 94.49 &  5.51 & 0.34 & 0.004 & 0.062 \\
                ND90-NH$_{2}$   & 94.03 &  5.41 & 0.56 & 0.006 & 0.103 \\
            \end{tabular}
        \end{ruledtabular}
    \end{table}

\begin{figure}[h]
        \includegraphics[width=0.5\textwidth]{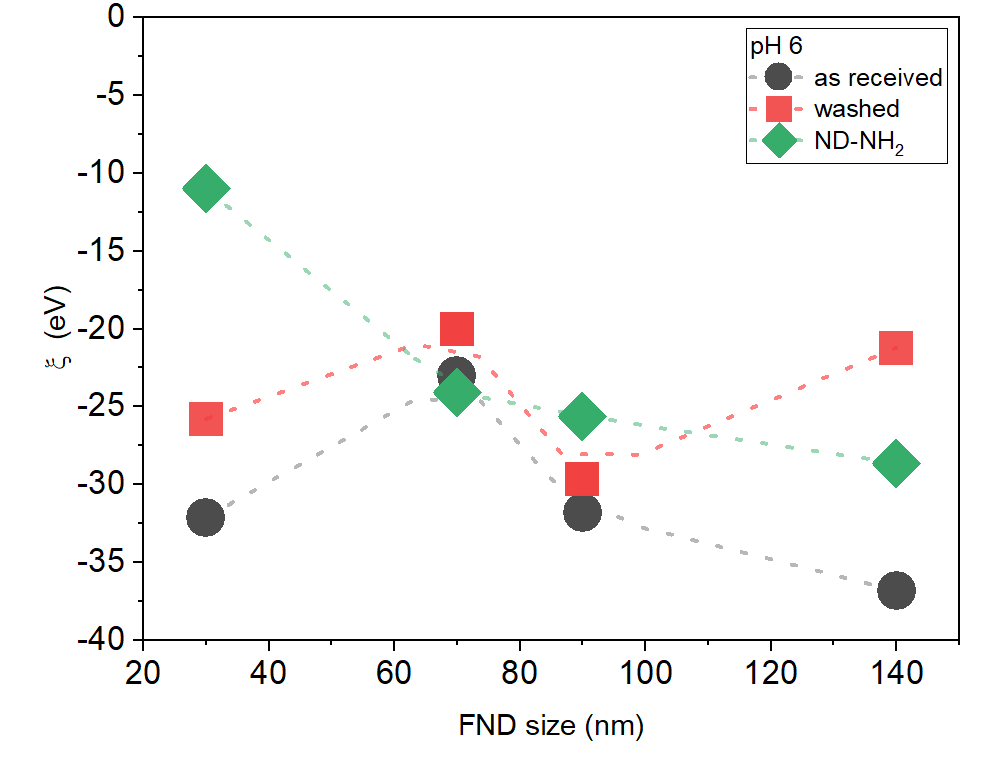}
        \caption{\label{fig:zeta} Zeta potential of as-received, washed, and amino-terminated (NH$_2$) nanodiamonds as a function of particle size. Measurements were conducted at pH~6.}
    \end{figure}

In order to further characterize the particle's surface modification, we have performed Zeta potential measurements. The measured zeta potential shifted toward more positive values following amino termination  (Figure~\ref{fig:zeta}), with a monotonic decrease as particle size increased. In contrast, the washed samples exhibited only a minimal increase compared to the as-received samples and displayed largely size-independent behavior. The zeta potential values of both the as-received and washed samples were consistent with previously reported values~\cite{Wilson2019, williams2010, apra2021}. While a consistent size-dependent zeta potential trend for FNDs has not been firmly established, carboxyl-terminated nanodiamonds typically exhibit negative zeta potentials around $-30$ to $-40$~mV.

Upon introduction of amino groups, a new type of basic functionality was created on the surface alongside the existing acidic carboxyl and hydroxyl groups. The measurements were done at pH 6. At this pH, carboxylic groups (pK$_a$~4.5) are predominantly deprotonated, whereas primary amines (pK$_a$~10--11) remain largely unprotonated, and therefore do not significantly contribute to a positive surface charge as NH$_3^+$ species. The nitrogen-containing functionalities present in our system cannot be regarded as simple, free primary amines. Instead, they are covalently bound to a chemically complex diamond surface. Their effective basicity and protonation behavior may therefore differ substantially from those of isolated amines in solution. 

The extent of amination varies across the different particle sizes. As particle size increases so does the surface complexity, partly due to the growing contribution of unreacted functional groups. This structural heterogeneity may also favor the formation of local zwitterionic configurations - analogous to those in amino acids - where spatially adjacent COO$^-$ and NH$_3^+$ groups could coexist, especially on larger, more complex surfaces. 

To support our findings with other measurements we compare the results of zeta potential and XPS analysis. The latter shows that the nitrogen surface coverage decreases with increasing particle size. Consequently, the relative contribution of nitrogen-containing surface species to the overall zeta potential diminishes for larger particles.

We conclude that the Hofmann degradation was successful according to the characterization of the termination of FNDs.

In addition to the accomplished termination of FNDs, a critical issue is the effects of it on the main properties of NV defect, as for the stabilization and for the contribution of the emission from NV($-$), because only NV($-$) emission from the total PL spectrum can be used for room-temperature quantum sensing. To monitor these properties after surface treatments, we characterize the FNDs using PL, ESR, and ODMR techniques. The ESR signals may provide insights about the electron spin environment around the NV centers in FNDs, including the surface spins (e.g., refs~\cite{Rondin2010, Barzegaramiriolya2025}).  

\begin{figure}[h]
        \includegraphics[width=0.5\textwidth]{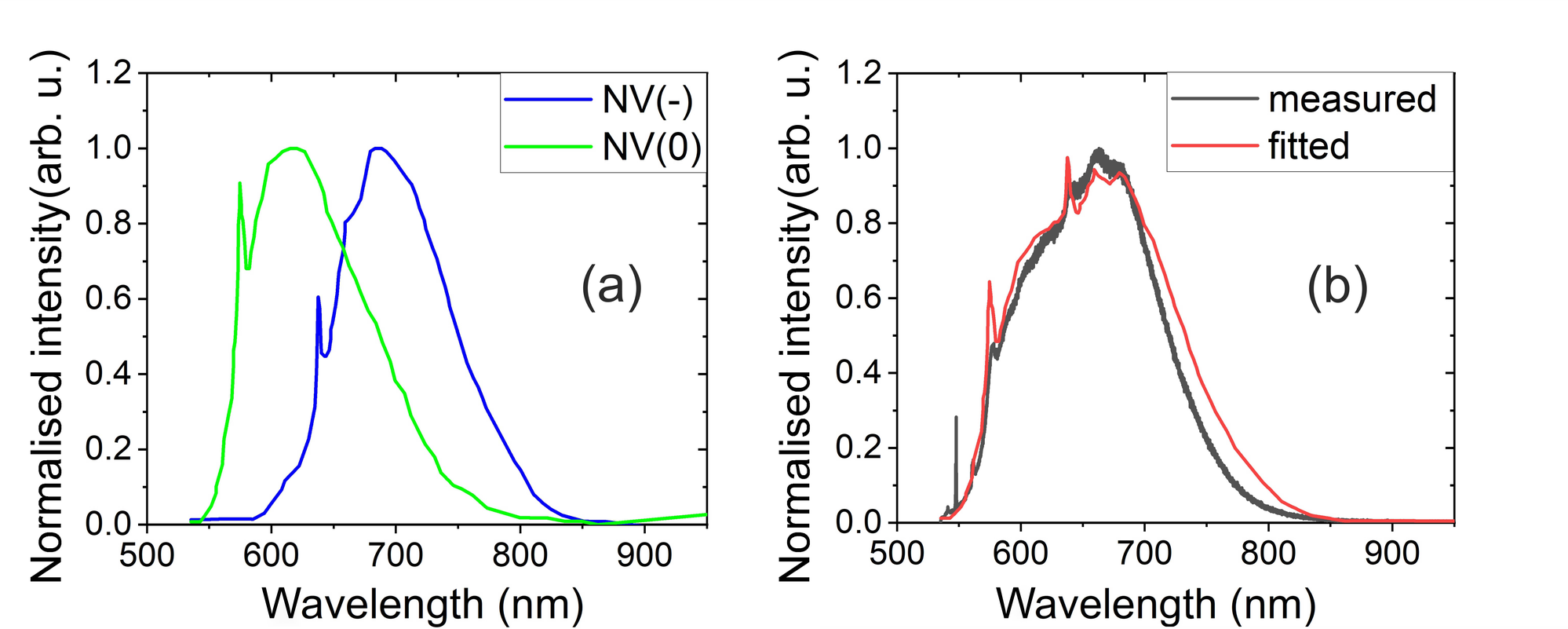}
        \caption{\label{fig:NV_PL} Photoluminescence spectrum analysis of FNDs. (a) Normalized photoluminescence spectra of NV($0$) and NV($-$) defects. (b) Photoluminescence spectrum of the as-received ND10 sample, where the fitted curve is the weighted superposition of NV($0$) and NV($-$) photoluminescence spectra.}
    \end{figure}

The PL signal is considered first. Under green illumination, the photoemission spectra of FNDs arise from the superposition of NV($-$) and NV($0$) emission spectra. This composite spectrum can be decomposed into a weighted sum of the known PL spectra of NV($-$) and NV($0$), as shown in Figure~\ref{fig:NV_PL}a. The relative contributions of the two charge states were determined following the method described in ref~\citenum{Rondin2010}, using the following procedure: (i) the measured PL spectra were first normalized to the maximum intensity of the NV($-$) phonon sideband, and (ii) the normalized spectra were then fitted by a linear combination of the reference PL spectra of NV($-$) and NV($0$), according to the equation,
    \begin{equation}
    I(\lambda) =  a_1 \times I_{\text{NV}(-)} (\lambda) + a_2 \times I_{\text{NV}(0)} (\lambda) \text{,} 
    \end{equation}
where $a_1$ and $a_2$ are the weighting factors. The ratio of fluorescence intensity produced by the NV($-$) and the total fluorescence intensity ($f_\text{NV}$) calculated as follows
\begin{equation}
    f_{\text{NV}(-)} = \frac{I_{\text{NV}(-)}}{I_{\text{NV}(-)}+I_{\text{NV}(0)}} \text{.}
\end{equation}
\rev{We emphasize that $f_{\text{NV}(-)}$ defined in this way is a purely optical quantity. It is extracted from the steady-state photoluminescence spectrum alone and involves no microwave manipulation of the NV spin, so it is set by the photodynamics of the surface and bulk defects and is independent of the magnetic-resonance protocol employed elsewhere in this work.}

    \begin{figure*}
        \includegraphics[width=0.9\textwidth]{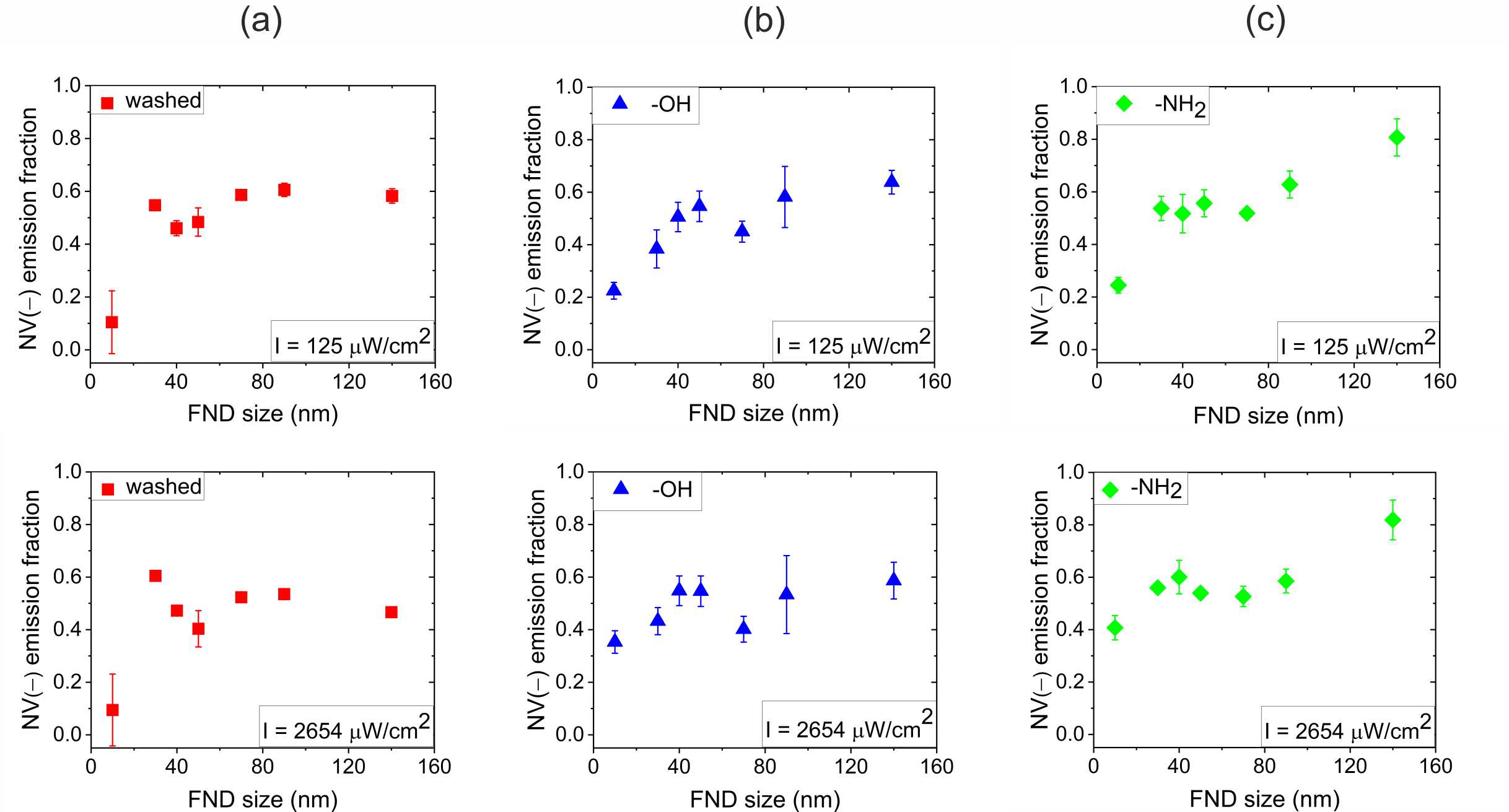}
        \caption{\label{fig:PL_FND_versus_density_size_2} Photoluminescence spectrum analysis of FNDs. NV($-$) emission fraction for the (a) washed, (b) $-$OH terminated, and (c) $-$NH$_2$ terminated FNDs as a function of the applied laser intensity for different FND sizes.}
    \end{figure*} 

A representative example of this deconvolution is shown for the as-received ND10 sample in Figure~\ref{fig:NV_PL}b. Typically, the goal in many FND-related studies is to enhance the NV$(-)$ emission, as this charge state exhibits optically detected magnetic resonance (ODMR) at room temperature. The observed $f_{\text{NV}(-)}$ values are influenced by both intrinsic and extrinsic factors affecting the NV centers. Notably, under continuous green laser illumination, NV centers cycle between the NV$(-)$ and NV$(0)$ charge states via two-photon ionization processes that involve excitation to the optical excited state~\cite{Gaebel2006,Gali2019}.  We note that understanding the microscopic mechanism of this process is still under intense research~\cite{Razinkovas2021, Wirtitsch2023, Thiering2024}. This is a nonlinear optical process that depends on the optical excitation power. Illumination may also ionize other defects in diamonds, such as the nitrogen donor, by ejecting an electron and vacancy clusters with ejecting holes, which depends on the doping concentration of nitrogen and the formation of NV defects via irradiation and annealing. We note that the latter process occurs for neutral vacancy clusters, which form due to band bending in near-surface regions~\cite{Neethirajan2023, Freire-Moschovitis2023}. Surface defects are typically acceptor-like,~\cite{Kaviani2014, Stacey2019, Dwyer2022, Chou2023, Dontschuk2023} that also eject holes after photoionization. As a consequence, the NV($-$) emission from the total emission is a photodynamics process, which depends on the excitation power, defect concentrations inside the FNDs, band bending from the surface toward the core of FNDs, and density of the surface defects on FNDs.

We plot the observed $f_{\text{NV}(-)}$ values as a function of size, surface termination, and laser illuminance in Figure~\ref{fig:PL_FND_versus_density_size_2}, where we show the data with the least and most intense laser powers (see Figure~S7). The $f_{\text{NV}(-)}$ of the as-received samples increased with particle size, following a power law and reaching a plateau at approximately 70~nm in diameter (this is shown in Figure~S6). This trend aligns with previous results reported for single NV($-$) centers~\cite{Rondin2010} and ensembles~\cite{Wilson2019}. 
For all other studied FNDs, the plateau occurred at smaller sizes: 40~nm for ND-OH and 20~nm for both washed and ND-NH$_2$ samples. This phenomenon can be understood by assuming that surface acceptor states are photoionizable with ejecting holes that would convert NV($-$) to NV($0$). The scattering $f_{\text{NV}(-)}$ data are comprehensible as the distances of NV defects from the diamond surface are not necessarily longer with larger FND size because of the irregular shape of FNDs. Another general trend is that increasing the power of the laser leads to larger $f_{\text{NV}(-)}$ values for the smallest FNDs (ND10) except for washed FNDs to which $f_{\text{NV}(-)}$ remains at $\sim 0.3$. The origin of this phenomenon is unclear as $f_{\text{NV}(-)}$ goes above 0.4 for as-received small FNDs (see Figure~S6); nevertheless, one may argue that intense laser irradiation may induce structural changes at the surface that remove acceptor states that have the most significant effect for the smallest FNDs. We do not have such evidence from FTIR studies that might be understood with the fact that these surface acceptor defects could be topographically protected and chemically stable at the interface of facets~\cite{Chou2023}. For the medium sized FNDs (50-90~nm) the $f_{\text{NV}(-)}$ values typically scatter between 0.5 and 0.6. The largest $f_{\text{NV}(-)}$ values occur for $-$NH$_2$ terminated ND140 FNDs at around 0.8 while it stays around 0.6 for as-received, washed or $-$OH terminated FNDs at the same size. This clearly demonstrates that the surface termination affects the properties of NV centers embedded in relatively large FNDs. FTIR analysis does not reveal any obvious reason for the remarkably high and laser power independent $f_{\text{NV}(-)}$ values for $-$NH$_2$ terminated ND140 FNDs as the FTIR spectra of $-$NH$_2$ FNDs exhibit similar features and characteristics in various sizes apart from small shifts (see Tables~S1--S7). The origin of this effect is instead revealed by the photoESR data discussed below (Table~\ref{tab:ESR_fit}): for the 140~nm particles, Hofmann degradation raises the relative weight of the P1 center while the broad $S=1/2$ signal of milling-induced surface dangling bonds is correspondingly diminished, indicating a net reduction of surface paramagnetic, acceptor-like defects. Because these surface acceptors are precisely the states whose photoionization ejects holes that convert NV($-$) into NV($0$), their suppression in ND140$-$NH$_2$ removes the dominant laser-power-dependent depletion channel, leaving an NV($-$) fraction that is both high and essentially independent of excitation power. This mechanism is specific to large particles: in smaller FNDs the NV centers sit closer to the surface and a larger surface-to-volume ratio sustains a residual acceptor density that the same chemistry cannot fully passivate, so the laser-power-independent plateau at $f_{\text{NV}(-)}\approx0.8$ does not develop. The amino termination therefore acts on the NV($-$) charge state indirectly, through the defect chemistry of the surface, rather than through a direct electron-affinity shift of the terminating groups.

Besides the $f_{\text{NV}(-)}$ ratio, also the photostability of NV($-$) is an essential point in FNDs that depends on the nearby acceptor and donor defects that could be paramagnetic; these paramagnetic species can be monitored by ESR spectroscopy. We performed electron spin resonance measurements on selected sample sizes (ND30, ND70, ND90, ND140), to characterize the paramagnetic centers in FNDs. 
The 
green illumination 
changes the charge state of defects affecting their spin states, thus we simulated this environment using photoESR studies, where the FND samples were irradiated by a 520~nm laser during the ESR measurements.  

\begin{figure}[h]
        \includegraphics[width=0.5\textwidth]{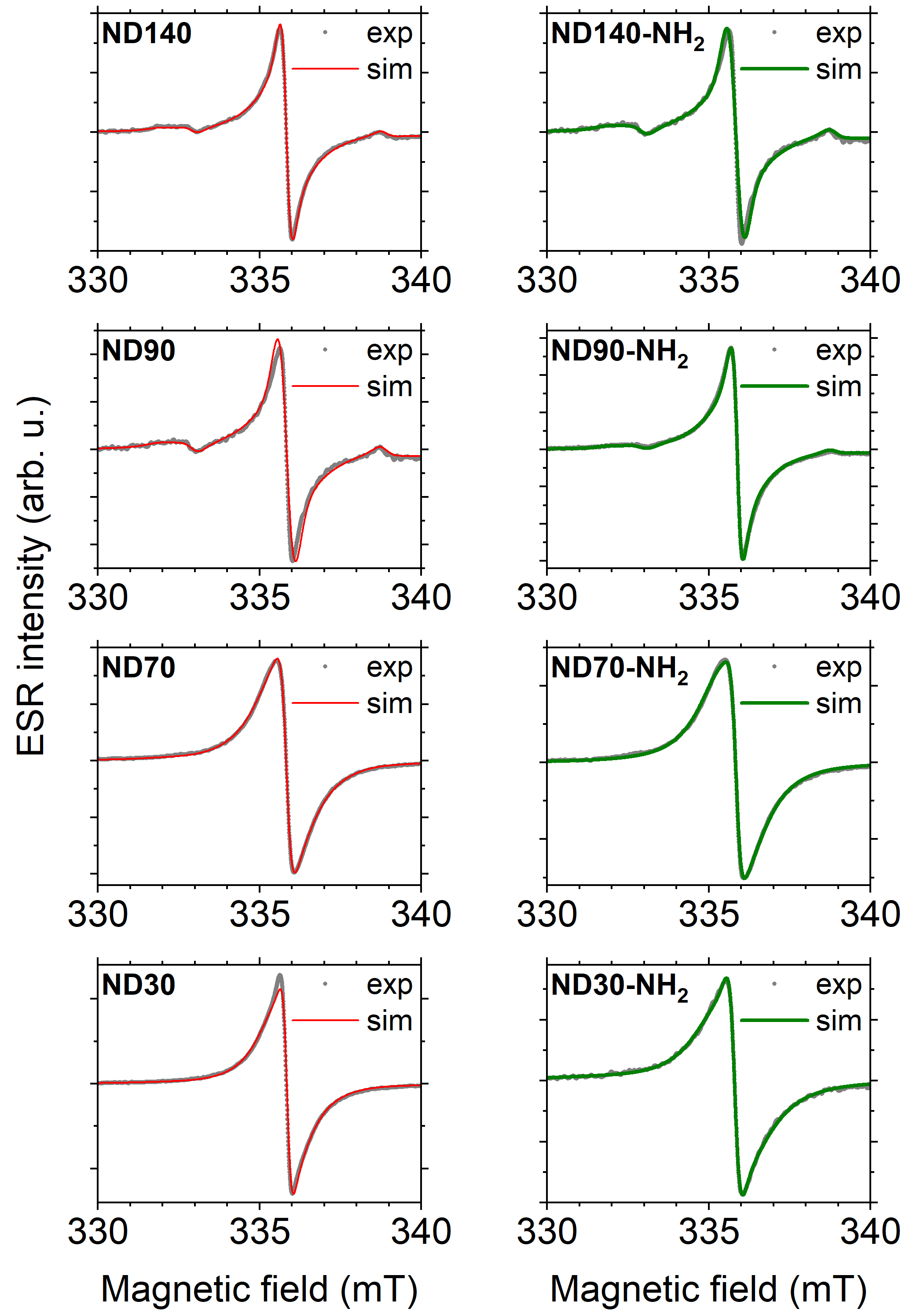}
        \caption{\label{fig:ESR_all-ND} ESR spectra and fitted plots for FNDs. Washed and $-$NH$_2$ terminated FNDs are shown in the left and right columns, respectively.}
    \end{figure}

Since the free amino-group is unstable in air, the measurements were performed in a liquid medium, \rev{1,4-dioxane} having a low dielectric constant of 2.25. The ESR spectra for the studied FNDs are plotted in Figure~\ref{fig:ESR_all-ND}. 

In agreement with previous findings~\cite{Shenderova2019, shames2015, panich2015}, the ESR spectra exhibit a central line that is the superposition of two Lorentzian signals: a broader signal attributed to carbon-inherited paramagnetic centers (primarily unpaired $S=1/2$ electron spins of dangling bonds formed during diamond milling), the P1 center that appears as a $^{14}$N hyperfine pattern broadened by the inhomogeneous strain field of nanodiamonds and a narrow $S=1/2$ ESR center. The origin of the narrow $S=1/2$ ESR center is not known. We hypothesize that it could be the ESR signal of paramagnetic vacancy clusters inside the nanodiamonds or close to the surface of nanodiamonds that are created by electron irradiation and annealing to mediate the conversion of substitutional nitrogen defects to NV defects. 

The main spin Hamiltonian parameters are listed in Table~\ref{tab:ESR_fit}. The $^{14}$N hyperfine pattern creates a visible signal outside of the broad central line at around 333 and 339~mT where the central line is dominated by the surface-related $S=1/2$ species in ND90 and ND140 samples. The signal of P1 center is masked by the other $S=1/2$ species for nanodiamonds smaller than 90~nm, although, it is weakly observable for ND70. Nevertheless, it can be concluded from the analysis of the ESR spectra of ND90 and ND140 samples that the signal of P1 center has higher relative intensity in amino-terminated FNDs than that in as-received FNDs. Since surface modification reactions affect only the outer moieties (thus altering the surface-related dangling bonds' concentration without impacting P1 centers), the observed increase in the relative intensity of the P1 center suggests a reduction of surface-related defects due to the Hofmann degradation process. Interestingly, the narrow $S=1/2$ ESR center does not show a clear trend upon Hofmann degradation of ND90 and ND140 samples. For ND90 samples, the relative concentration of the narrow $S=1/2$ ESR center decreases after Hofmann degradation while it increases for ND140 samples. This result implies that if the narrow $S=1/2$ ESR defects reside close to the surface of nanodiamonds they are not directly affected by the surface modifications. This might be a complex interplay of various effects: as the number of dangling bonds decreases at the surface upon Hofmann degradation process the band bending may incline downward due to $-$NH$_2$ groups but the near-surface substitutional nitrogen donors could easily transfer their electrons toward the acceptor-like near surface vacancies~\cite{Peng2019} that becomes paramagnetic. With assuming the charge stabilization of the paramagnetic defects near the NV center in ND140 samples after Hofmann degradation process, the outstanding $f_{\text{NV}(-)}$ might be explained that does not occur for smaller FNDs.

\rev{It should be emphasized that the spectra in Figure~\ref{fig:ESR_all-ND} report exclusively on $S=1/2$ species, and that the NV($-$) centers embedded in the very same particles contribute no observable signal. This is expected rather than surprising. NV($-$) has an $S=1$ ground state whose zero-field splitting, $D=2.87$~GHz, corresponds to $\approx102$~mT in field units at $g\approx2$, i.e., 3 orders of magnitude larger than the line widths of the $S=1/2$ signals analyzed above. In a randomly oriented nanodiamond powder the allowed $\Delta m_S=\pm1$ transitions of such a triplet are consequently not concentrated near $g\approx2$ but are dispersed into a broad powder pattern spanning roughly $\pm102$~mT about the free-electron field, whose most intense (perpendicular) turning points already lie $\approx51$~mT away from it, far outside the narrow field window in which the P1 hyperfine lines and the surface $S=1/2$ signal appear. The pattern is smeared further by the particle-to-particle distribution of $D$ and $E$ produced by the inhomogeneous strain of the nanodiamonds, which the ODMR data below quantify directly (Table~\ref{tab:ODMR}). Together with the low NV content of these samples ($\leq3$~ppm, i.e., from one or two to a few hundred centers per particle, Table~\ref{tab:FND}) compared with the P1 and dangling-bond densities, this places the NV contribution below the detection limit of the conventional field-swept experiment.}

\rev{This is precisely why the NV center must be addressed optically instead. The spin-selective fluorescence of NV($-$) supplies both the sensitivity and the defect selectivity that field-swept ESR lacks here, and the cw-ODMR spectra presented in the following section resolve the $D\pm E$ structure of the very same centers that remain invisible in Figure~\ref{fig:ESR_all-ND}. The two techniques are therefore complementary rather than redundant: ESR characterizes the paramagnetic \textit{bath}, that is, the P1 centers and the milling-induced surface dangling bonds that generate the magnetic and electric-field noise experienced by the NV centers, whereas ODMR probes the NV \textit{sensor} itself. We make explicit use of this complementarity below, where the size and termination dependence of the $E$ parameter measured by ODMR is shown to follow the weight of the broad $S=1/2$ surface signal measured by ESR, thereby linking the spin properties of the NV sensor directly to the surface defect chemistry.}

    \begin{table*}[t]
        \caption{\label{tab:ESR_fit}Simulated electron spin resonance spectra with using three paramagnetic defects. The main spin Hamiltonian parameters are listed where the $g$ factor of the P1 center is fixed at the experimental data~\cite{smith1959, shames2015, panich2015}. The ESR parameters determined for the P1 centers by simulation are $g_{\mathrm{iso}} = 2.0024$, $A_{zz} = 115$~MHz and $A_{xx} = A_{yy} = 82$~MHz. The individual line widths of $\Delta H_{\mathrm{pp}}^{\mathrm{L}}$ with Lorentzian function and the strain applied in the fitting procedure of the hyperfine interaction are shown in MHz unit. The  $A_{\mathrm{strain}}$ values are for $A_{zz}$ and $A_{xx} = A_{yy}$, respectively. The estimated relative uncertainty of the fitted line widths ($\Delta H_{\mathrm{pp}}^{\mathrm{L}}$), strain values ($A_{\mathrm{strain}}$), and relative weights is about $\pm3\%$, reflecting the reproducibility of the EasySpin simulations over repeated measurements; the $g$-factors are quoted with their absolute fit uncertainties in parentheses [e.g., $2.0030(2)$ denotes $2.0030\pm0.0002$].}
        \begin{ruledtabular}
            \begin{tabular}{l|cccc ccc ccc}
             & \multicolumn{4}{c}{P1 center ($S=1/2$)} & \multicolumn{3}{c}{Broad center ($S=1/2$)} & \multicolumn{3}{c}{Narrow center ($S=1/2$)} \\
            Sample & $g$ & $\Delta H_{\mathrm{pp}}^{\mathrm{L}}$ & $A_{\mathrm{strain}} $ & Weight & 
                     $g$ & $\Delta H_{\mathrm{pp}}^{\mathrm{L}}$ & Weight &
                     $g$ & $\Delta H_{\mathrm{pp}}^{\mathrm{L}}$ & Weight \\
            \hline
            ND30 
             & - & - & - & 0.0000
             & 2.0030(2) & 1.002 & 0.8749
             & 2.0029(2) & 0.006 & 0.1251 \\
            ND30-NH$_2$
             & - & - & - & 0.0000
             & 2.0030(2) & 1.187 & 0.9231
             & 2.0029(2) & 0.349 & 0.0768 \\
            ND70 
             & 2.00240   & 0.710 & 10.68, 7.03 & 0.0022(18)
             & 2.0030(2) & 1.174 & 0.9103
             & 2.0028(2) & 0.388 & 0.0875 \\
            ND70-NH$_2$
             & 2.00240 & 0.775 & 10.68, 5.54 & 0.0024(16)
             & 2.0030(2) & 1.61 & 0.9109
             & 2.0028(2) & 0.712 & 0.0867 \\
            ND90
             & 2.00240 & 0.691 & 10.68, 7.35 & 0.0257
             & 2.0028(2) & 1.103 & 0.7665
             & 2.0027(2) & 0.306 & 0.2078 \\
            ND90-NH$_2$
             & 2.00240 & 0.717 & 10.68, 5.21 & 0.0768
             & 2.0030(2) & 1.261 & 0.7473
             & 2.0029(2) & 0.348 & 0.1760 \\
            ND140
             & 2.00240 & 0.419 & 10.68, 3.01 & 0.1074
             & 2.0030(2) & 1.308 & 0.7503
             & 2.0029(2) & 0.356 & 0.1423 \\
            ND140-NH$_2$
             & 2.00240 & 0.481 & 10.68, 1.65 & 0.1185
             & 2.0030(2) & 1.457 & 0.7265
             & 2.0029(2) & 0.362 & 0.1550 \\
            \end{tabular}
        \end{ruledtabular}
    \end{table*}

\begin{figure}
        \includegraphics[width=0.5\textwidth]{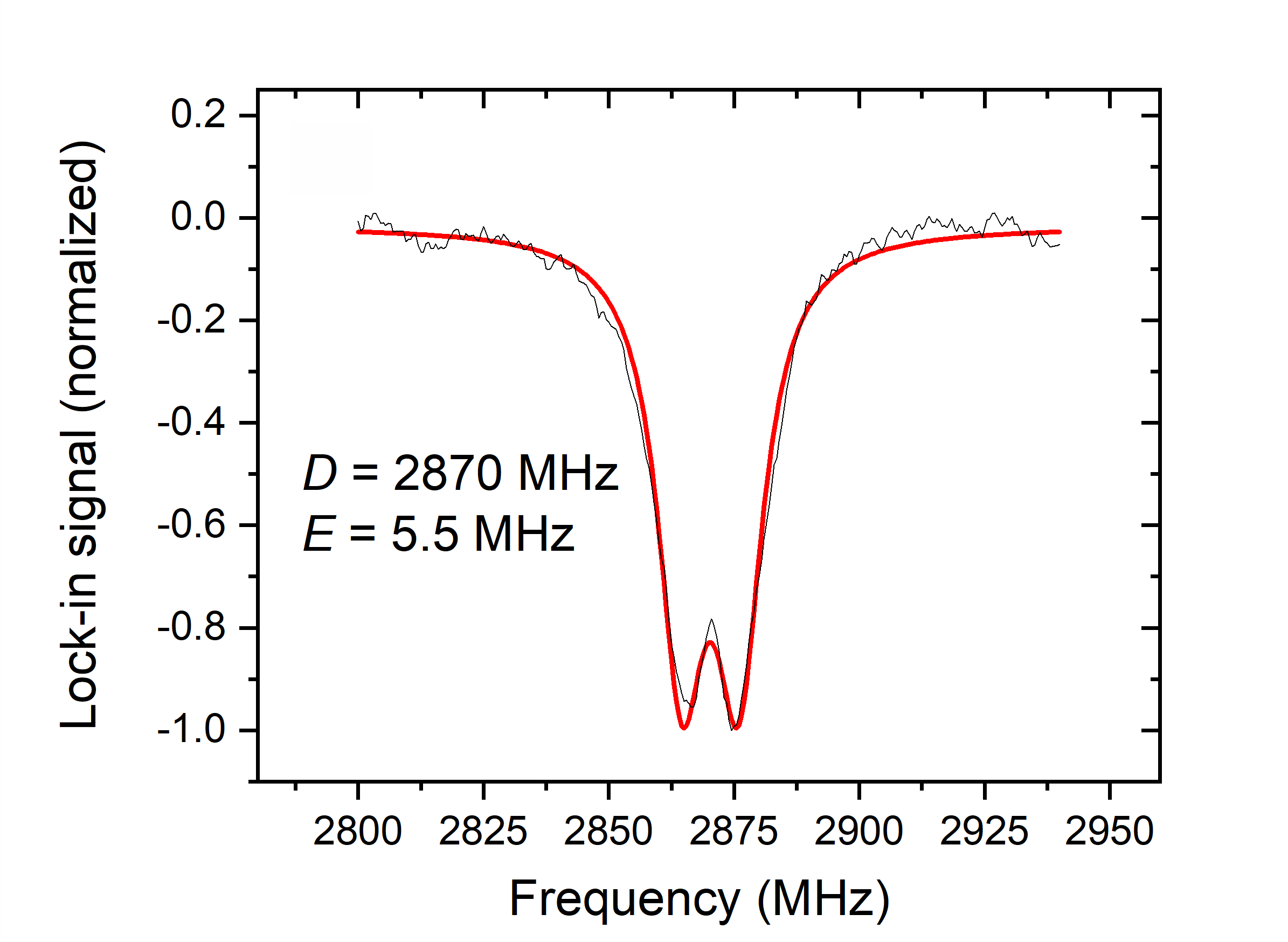}
         \caption{\label{fig:ODMR} CW-ODMR results on ND50-OH. 
         Red curves are simulated data.}
    \end{figure}
\begin{figure}[h]
        \includegraphics[width=0.5\textwidth]{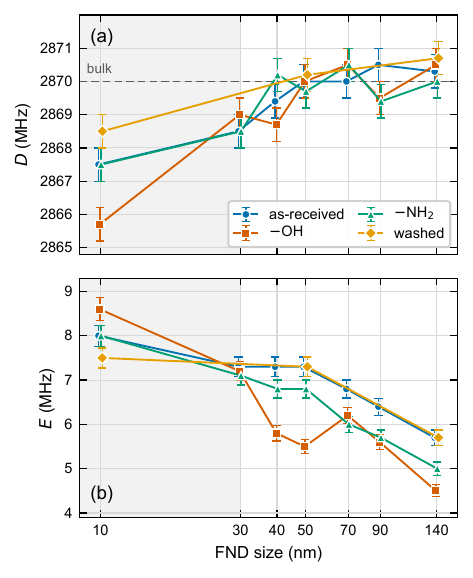}
        \caption{\label{fig:ZFS}\rev{Zero-field splitting parameters of the NV($-$) ensembles as a function of FND size, for all four surface terminations: (a) the axial parameter $D$ and (b) the symmetry-breaking parameter $E$. The plotted values are those listed in Table~\ref{tab:ODMR}, and the error bars are the fit uncertainties quoted there ($\pm0.5$~MHz for $D$, $\pm3\%$ for $E$). The dashed line in (a) marks the bulk-diamond value $D=2870$~MHz; the shaded region indicates the $\leq30$~nm range over which $D$ departs from it, which we attribute to static strain. The two panels differ in character: $E$ decreases monotonically with size for every termination, whereas $D$ is constant within the fit uncertainty above 30~nm. Lines connecting the symbols are guides to the eye.}}
    \end{figure}

The quantum sensor operation of FNDs is based on the ODMR signal of the NV($-$) defects. The ODMR technique combines photoluminescence with electron spin resonance. To complete our study of the FNDs, we record the ODMR signals for the ensemble of particles. The intensity of fluorescence was relatively weak for the smallest FNDs (30~nm) in our investigation, consequently,  we picked up the brightest spots that we found in the confocal map of the samples. A typical cw-ODMR signal is plotted for the representative ND50-OH FNDs in Figure~\ref{fig:ODMR}. The thin black curve is the recorded ODMR data, whereas the red curve is the simulation of the ODMR signal with the use of the spin-Hamiltonian of NV($-$) defect with allowing inhomogeneous broadening. Apparently, a double negative peak arises, which is a signature of breaking the $C_{3v}$ symmetry of the defect. The small valley between the two peaks is the $D$ parameter, whereas the two peaks are separated by $2\times E$. We recorded the ODMR signal for every type of FNDs considered in our study and fit the results to extract the $D$ and $E$ parameters that we list in Table~\ref{tab:ODMR}. We find a general trend that the $E$ parameters shift from $\sim8$~MHz toward $\sim5$~MHz as we increase the size of FNDs from 30 to 140~nm, independently of the surface termination. The size dependence of the observed $D$ parameters differs: it shows up $\sim2867$~MHz in 10~nm sized FNDs, then it increases to $\sim2870$~MHz for larger FNDs (40~nm), after that the observed $D$ parameters fluctuate around this value for larger FNDs. \rev{Both trends are plotted in Figure~\ref{fig:ZFS}, which displays the data of Table~\ref{tab:ODMR} graphically. The figure makes the qualitative difference between the two parameters immediately apparent: $E$ falls monotonically with size for every surface termination, whereas $D$ is displaced from its bulk value only in the two smallest particle sizes and is otherwise constant within the fit uncertainty.}

\begin{table*}
        \caption{Magnetic resonance parameters for various investigated FNDs. The estimated fit uncertainty of the $E$ parameter is about $\pm3\%$ ($\sim\pm0.2$~MHz). For the $D$ parameter, a relative figure is not meaningful, as its line centre near 2870~MHz is determined far more precisely than $3\%$ would imply; we therefore quote an absolute fit uncertainty of $\pm0.5$~MHz, which renders the $\sim3$~MHz shift observed in the smallest ($\leq30$~nm) FNDs statistically significant.}
        \label{tab:ODMR}
        \begin{ruledtabular}
             \begin{tabular}{c|cc|cc|cc|cc}
               NDx  & \multicolumn{2}{c|}{as-received} &\multicolumn{2}{c|}{$-$OH}  &\multicolumn{2}{c|}{$-$NH$_2$} &\multicolumn{2}{c}{washed} \\ \hline
                $d$& $D$ & $E$  & $D$ & $E$ & $D$ & $E$ &  $D$ & $E$  \\
                (nm)& (MHz) & (MHz) & (MHz) & (MHz) & (MHz) & (MHz) &  (MHz) & (MHz) \\ \hline
                10   &   2867.5 & 8.0       & 2865.7 & 8.6           & 2867.5 & 8.0     &   2868.5 & 7.5       \\ 
                30       & 2868.5 & 7.3        & 2869.0 & 7.2    & 2868.5   & 7.1        &-&-\\ 
                40       & 2869.4 & 7.3      &  2868.7 & 5.8      & 2870.2 & 6.8    &-&- \\ 
                50       & 2870.0   & 7.3    & 2870.0 & 5.5        & 2869.7  & 6.8     &   2870.2 & 7.3     \\ 
                70       & 2870.0   & 6.8        & 2870.5 & 6.2    & 2870.5 & 6.0     &-&- \\
                90       & 2870.5 & 6.4    & 2869.5   & 5.6       &  2869.4 & 5.7    &-&-  \\ 
                140      & 2870.3 & 5.7     & 2870.5   & 4.5     & 2870.0 & 5.0    &   2870.7 & 5.7  \\
            \end{tabular}
        \end{ruledtabular}
    \end{table*}

There could be two reasons for the symmetry-breaking $E$ splitting in the ODMR spectrum: static strain and (fluctuating) electric fields. Theoretical analysis and simulations show that two fields dissimilarly affect the ODMR spectrum on an ensemble of NV defects: static strain both shifts the $D$ parameter and creates an $E$ splitting, whereas the electric fields do not change the $D$ parameter but only create $E$ splitting~\cite{Mittiga2018}. The data in Table~\ref{tab:ODMR} implies that the static strain affects the spin levels of the electronic ground state in 10 and 30~nm sized FNDs, whereas it has a minor effect in larger FNDs as the $D$ parameters do not shift in the latter. According to a recent \textit{ab initio} study, the $E$ parameter fast converges below 5~MHz with increasing the distance between NV($-$) and the (100) facet of the diamond surface~\cite{Pershin2025}. For larger nanodiamonds, it is more likely that NV centers reside farther from the diamond surface than 2~nm, thus our conclusion is well supported by this study. The size-dependent change in the $E$ parameter of the spin levels in the NV centers' ground state embedded in larger ($>30$~nm sized) FNDs should be associated with the electric field noise. The electric field noise around the NV centers is typically caused by charging the parasitic defects of diamond with illumination that is used to spin-polarize optically or read out the spin state. The source of the parasitic defects could be the substitutional nitrogen donor defects, acceptor-like vacancy-cluster defects around the NV center, or acceptor-like near-surface or surface defects. The distance between the surface defects and the NV center increases in larger FNDs which reduces the value of the $E$ parameter. This explanation is consistent with the interpretation of the ESR data, where the density of the broad $S=1/2$ ESR centers associated with surface dangling bonds decreases either with increasing size of FNDs or going from the as-received to amino-terminated samples with a given size of FNDs, and the respective $E$ parameters follow the same trend.

\section{Conclusion}
\label{sec:summary}
We studied the fundamental properties of FNDs embedding NV centers, including particle shape, the vibrational spectrum of surface groups, the PL spectrum of NV defects, and paramagnetic defects under illumination, which are also used to detect the ODMR signal from NV centers. In particular, we applied the Hofmann degradation as an innovative, wet-chemical approach to achieve amino termination of the FNDs. The successful termination was confirmed by FTIR, while the XPS confirmed the presence of nitrogen on the surfaces.

Beyond this materials characterization, two findings constitute the central, novel contribution of this work. First, by resolving the NV($-$) zero-field splitting across the full 10$-$140~nm size series, we establish that the symmetry-breaking $E$ parameter decreases monotonically with increasing FND size for every surface termination, whereas the axial $D$ parameter departs from its bulk value only in the smallest ($\leq30$~nm) particles. Following the strain-versus-electric-field framework of ref~\citenum{Mittiga2018} and the \textit{ab initio} convergence of the $E$ parameter with NV-to-surface distance~\cite{Pershin2025}, this disentangles a static-strain contribution that dominates the spin levels below $\sim30$~nm from a fluctuating electric-field contribution, set by photocharged parasitic defects, whose magnitude scales with the NV-to-surface distance and therefore with particle size. This assignment is independently corroborated by the ESR data, in which the density of the broad $S=1/2$ surface-dangling-bond signal follows the same size and termination trends as the $E$ parameter. Second, ESR measurements proved that the Hofmann procedure did not induce further surface defects, as is usually the case during plasma treatment for amino termination; instead, the photoESR data reveal a net degradation of surface paramagnetic defects, evidenced by the increased relative weight of the P1 center. We identify this suppression of surface acceptor-like defects as the origin of the remarkably high and laser-power-independent NV($-$) fraction ($f_{\text{NV}(-)}\approx0.8$) observed uniquely in 140~nm amino-terminated FNDs, where the NV centers lie far enough from the surface for the passivation to be effective.

We conclude that wet-chemical Hofmann amino termination simultaneously preserves a near-bulk, strain-free NV environment and maximizes the NV($-$) content in 140~nm FNDs, while providing amino groups for direct biomolecule attachment. The combination makes these amino-terminated FNDs a promising platform for room-temperature quantum sensing in chemical and biological settings.




\begin{acknowledgement}
The authors thank Prof.\ Ferenc Simon for providing access to the ESR instrument. The work was supported by the Quantum Information National Laboratory sponsored by National Research, Development and Innovation Fund (NKFIH) grant no.\ 2022-2.1.1-NL-2022-00004, K137852, K149457, TKP2021-EGA-02, TKP2021-NVA-02, and grant no.\ VEKOP-2.3.3-15-2016-00002 of the European Structural and Investment Funds.
The research reported in this paper and carried out at Wigner Research Centre for Physics is supported by the infrastructure of the Hungarian Academy of Sciences.
\end{acknowledgement}

\section*{Author Contributions}
Conceptualization, N.J., D.B. and A.G.; methodology, N.J., O.K., Zs.Cz., D.B. and A.G.; investigation, N.J., V.V., O.K., Zs.Cz., Sz.Cz., A.Cs., S.K.; project administration, A.G.; resources, A.G.; supervision, A.G.; writing---original draft, N.J., D.B.\ and A.G. All authors have read and agreed to the published version of the manuscript.


\begin{suppinfo}
The following file is available free of charge.
\begin{itemize}
  \item \textbf{Supporting Information (PDF)}: morphological and structural characterization by SEM, SEM-EDS, and HRTEM-EDS (Figures~S1--S3); size-resolved FTIR spectra of all studied FND sizes (Figure~S4) together with detailed vibrational band assignments (Tables~S1--S7); supporting C~1\textit{s} and N~1\textit{s} XPS spectra (Figure~S5); and additional power-dependent NV($-$) emission-fraction data (Figures~S6 and S7).
\end{itemize}
\end{suppinfo}

\providecommand{\latin}[1]{#1}
\makeatletter
\providecommand{\doi}
  {\begingroup\let\do\@makeother\dospecials
  \catcode`\{=1 \catcode`\}=2 \doi@aux}
\providecommand{\doi@aux}[1]{\endgroup\texttt{#1}}
\makeatother
\providecommand*\mcitethebibliography{\thebibliography}
\csname @ifundefined\endcsname{endmcitethebibliography}
  {\let\endmcitethebibliography\endthebibliography}{}

\clearpage

\begin{center}
{\large\bfseries Supporting Information}\\[4pt]
{\itshape This document contains supporting morphological, structural, vibrational, X-ray photoelectron, and photoluminescence data for the main text.}
\end{center}

\setcounter{section}{0}
\renewcommand{\thesection}{S\arabic{section}}
\renewcommand{\thefigure}{S\arabic{figure}}
\renewcommand{\thetable}{S\arabic{table}}
\setcounter{figure}{0}
\setcounter{table}{0}

\section{Analysis by SEM, HRTEM, and EDS}
\label{app:SEM}
\begin{figure}[b]  
    \includegraphics[width=0.5\textwidth]{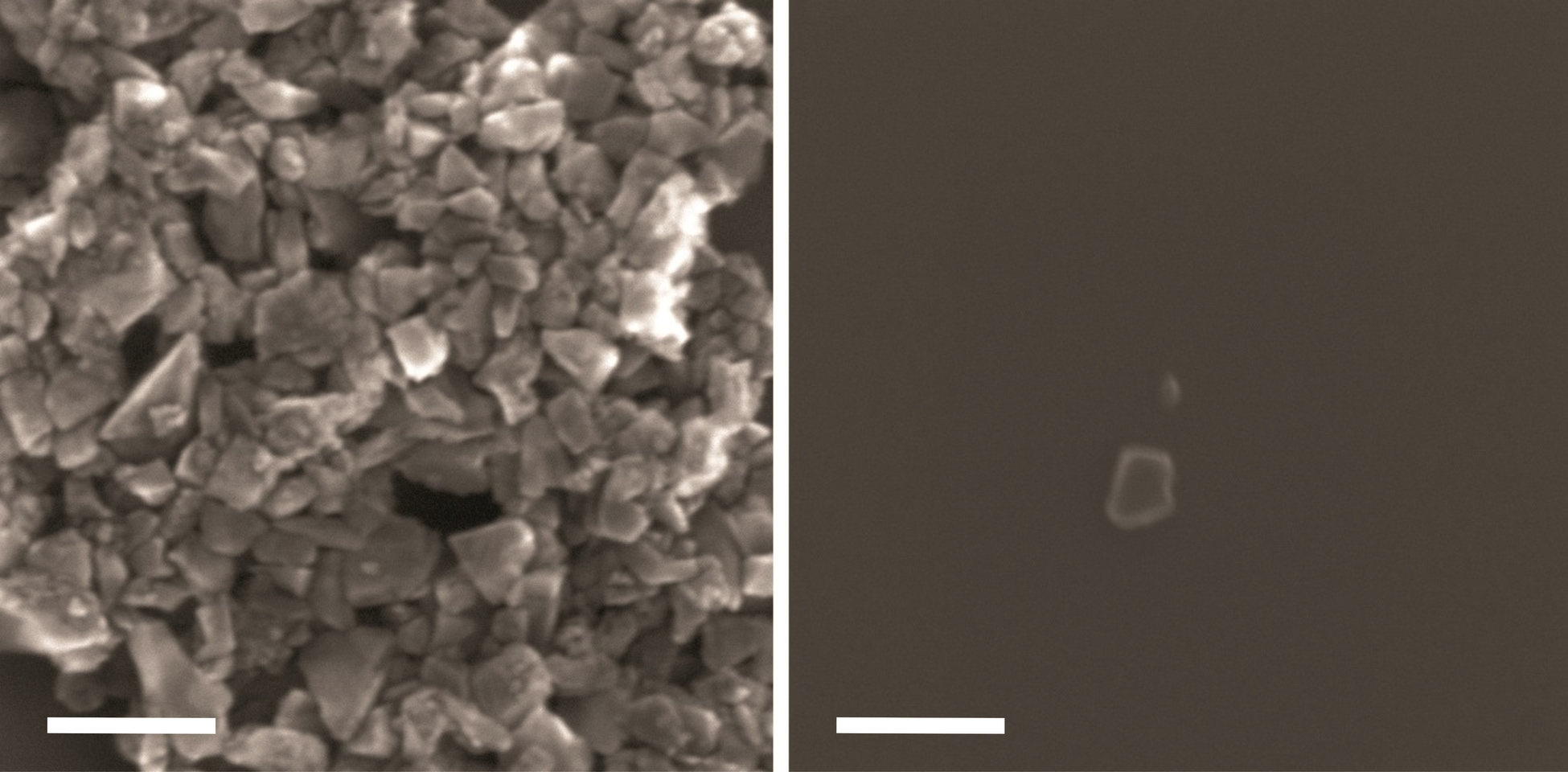}
    \caption{\label{fig:SEM}SEM image of ND90 ensemble and individual particles. The white line corresponds to 200~nm.}
    \end{figure}

\begin{figure}[b!]  
    \includegraphics[width=0.45\textwidth]{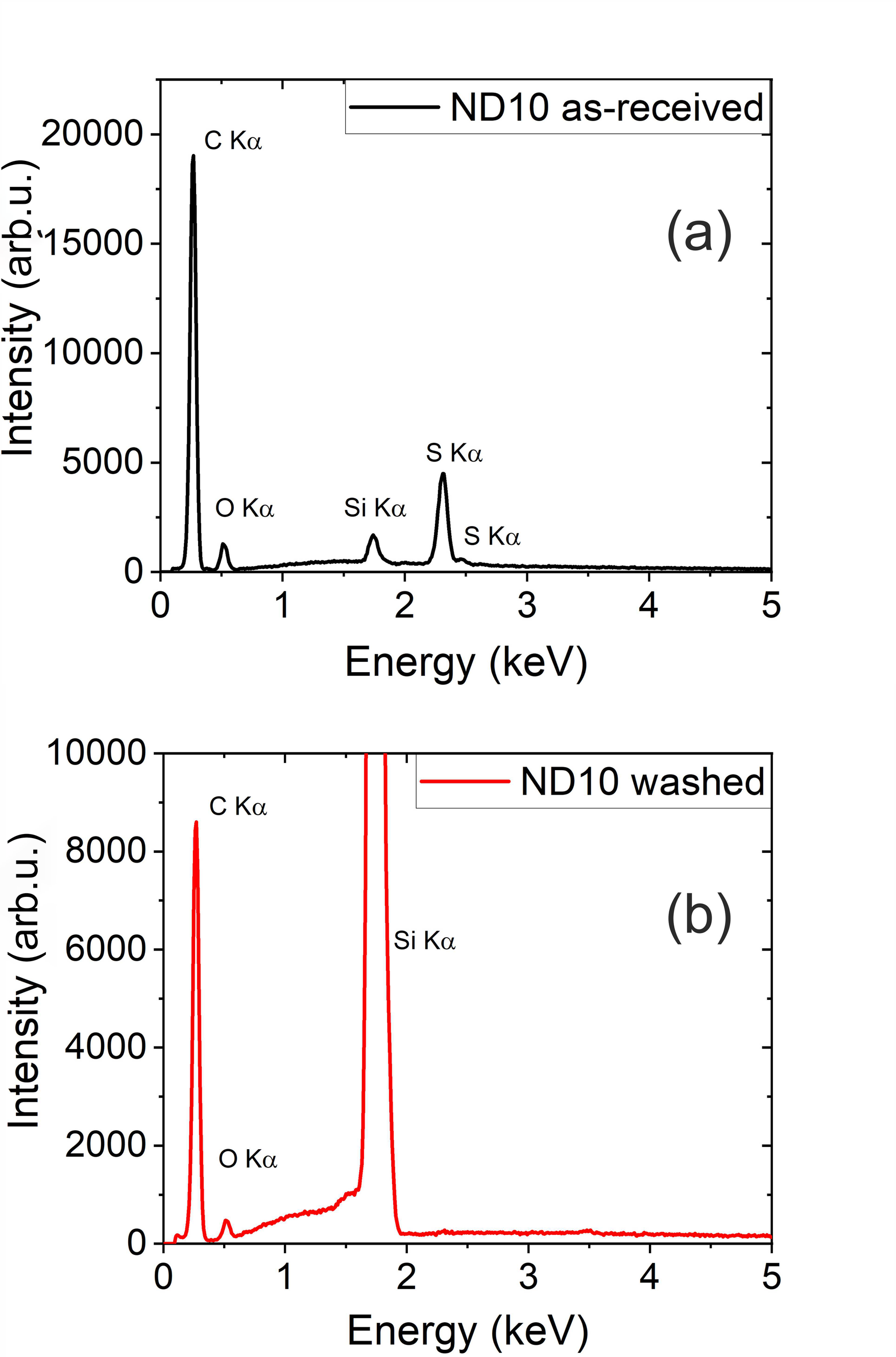}
    \caption{\label{fig:EDS} EDS spectra of (a) ND10 as-received and (b) ND10 washed particles. }
    \end{figure}

\begin{figure*}
    \centering
    \includegraphics[width=0.9\textwidth]{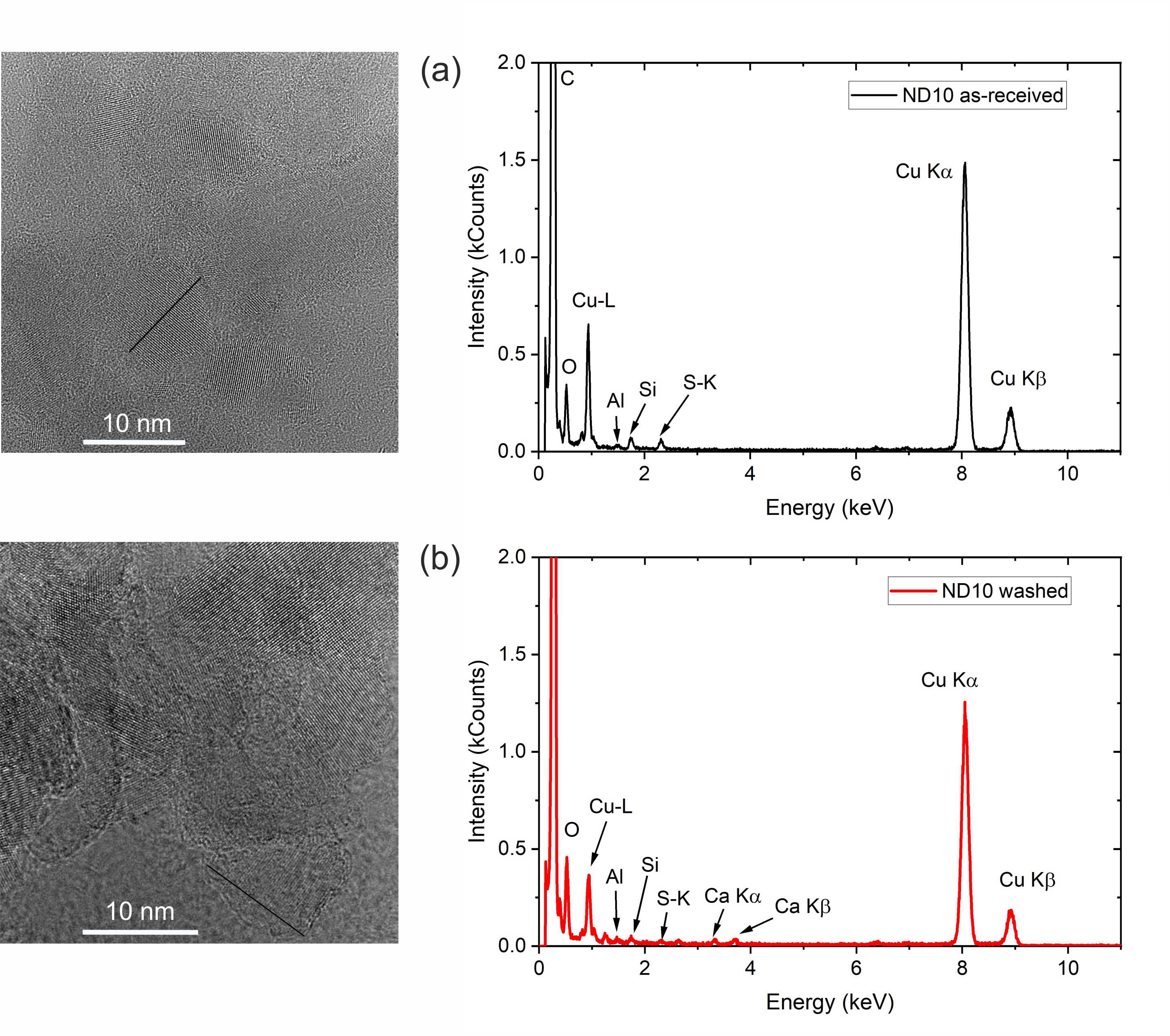}
    \caption{\label{fig:TEM} HRTEM image and TEM-EDS analysis of (a) ND10 as-received and (b) ND10 washed particles. The scale corresponds to 10~nm. The black line indicates the measurement of the size of the nanodiamond particle.}
    \end{figure*}

The SEM images clearly show the non-spherical shape of the FNDs in Fig.~\ref{fig:SEM}. Fig.~\ref{fig:EDS} presents the EDS spectra of ND10 as-received and washed samples as obtained by SEM. The presence of sulfur is evident in the as-received samples. This observation was also proven by the EDS analysis performed in TEM (Fig.~\ref{fig:TEM}).

\FloatBarrier
\section{FTIR supporting data}
\label{app:FTIR}

\begin{figure*}
\includegraphics[width=0.7\textwidth]{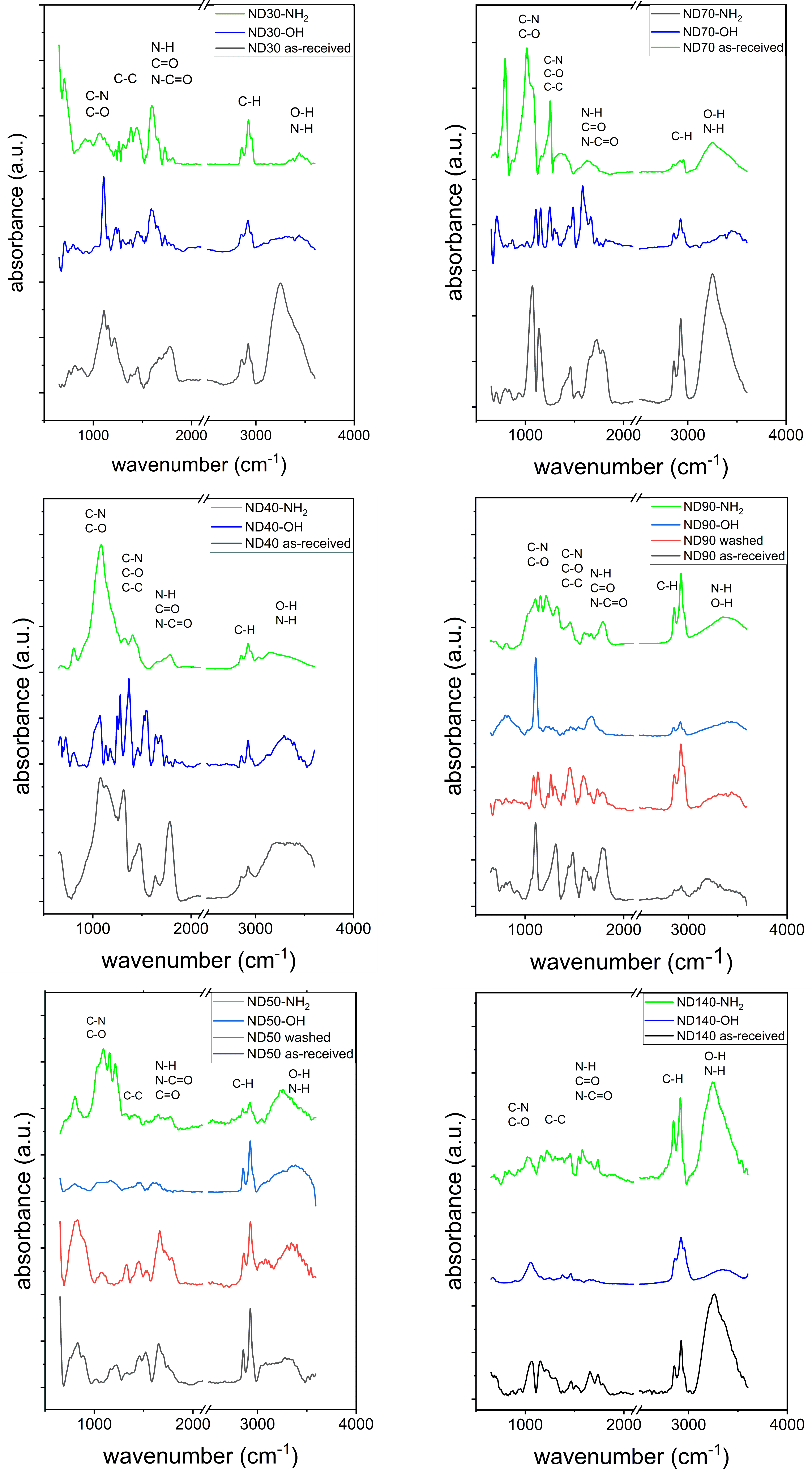}
\caption{\label{fig:IR-140ND}Infrared vibration spectra of the as-received, $-$OH and $-$NH$_{2}$ terminated FNDs with average sizes of 30, 40, 50, 70, 90 and 140~nm}
\end{figure*}

In Tables~\ref{tab:FTIR-ND10}–\ref{tab:FTIR-ND140}, the vibrational bands identified in the FTIR spectra (shown in Figure~2 of the main text and Figure~\ref{fig:IR-140ND}) are summarized together with their corresponding surface functional group assignments, based on standard FTIR band assignments. FTIR measurements were performed on as-received, washed, hydroxyl-terminated, and amino-terminated samples of the ND10, ND50, and ND90 FNDs. For the ND30, ND40, ND70, and ND140 FNDs, the washed samples were not examined. The FTIR spectra of ND40 closely resembled those of ND30; therefore, the ND40 spectra are not presented in detail.  Note that the functional group $-$SO$_3$X appears in several places in the tables, where "X" stands for H (sulfonic acid group) or Na (sulfonate salt). In the table we mainly highlight those peaks, which are typical fingerprints for the different terminations. 

\begin{table*}[t]
  \caption{\label{tab:FTIR-ND10}FTIR data measured on the ND10 FNDs with various surface terminations. The observed vibration bands, the associated functional groups, and the respective vibration modes are given.}
  \begin{ruledtabular}
  \begin{tabular}{l l l}
    band (cm$^{-1}$) & functional groups & vibration modes \\ \colrule
    \multicolumn{3}{l}{ND10 as-received} \\
    3590--3130 & \parbox[t]{3cm}{COOH, OH} &
                 \parbox[t]{11cm}{O--H stretching} \\
    2960--2856 & \parbox[t]{3cm}{CH$_3$, CH$_2$} &
                 \parbox[t]{11cm}{C--H stretching} \\
    1850--1660 & \parbox[t]{3cm}{COOH, CO, CHO, C=C} &
                 \parbox[t]{11cm}{C=O from carboxyl group (monomer, dimer), acid-anhydride (symmetric and asymmetric), H bond in chelate bond, C=C stretching} \\
    1440       & \parbox[t]{3cm}{CH$_3$, C--C} &
                 \parbox[t]{11cm}{bending mode, perpendicular to the plane, C--C skeletal} \\
    1270--1180 & \parbox[t]{3cm}{SO$_3$X, C--O--C} &
                 \parbox[t]{11cm}{from sulfonate, symmetric and asymmetric C--O--C stretching} \\
    1180--1000 & \parbox[t]{3cm}{OH} &
                 \parbox[t]{11cm}{C--O stretching modes (primary, secondary, tertiary)} \\
    \colrule
    \multicolumn{3}{l}{ND10 $-$OH terminated} \\
    3600--3120 & \parbox[t]{3cm}{OH} &
                 \parbox[t]{11cm}{O--H stretching} \\
    2953--2850 & \parbox[t]{3cm}{CH$_3$, CH$_2$} &
                 \parbox[t]{11cm}{C--H stretching} \\
    1660       & \parbox[t]{3cm}{C=O} &
                 \parbox[t]{11cm}{C=O from unreacted COOH group, C=C stretching} \\
    1463       & \parbox[t]{3cm}{CH$_3$} &
                 \parbox[t]{11cm}{bending mode, perpendicular to the plane} \\
    1240--1000 & \parbox[t]{3cm}{C--O, C--C} &
                 \parbox[t]{11cm}{C--O stretching modes (primary, secondary, tertiary), skeletal vibration} \\
    \colrule
    \multicolumn{3}{l}{ND10 $-$NH$_2$ terminated} \\
    3320       & \parbox[t]{3cm}{NH$_2$, NH} &
                 \parbox[t]{11cm}{N--H stretching} \\
    2960--2855 & \parbox[t]{3cm}{CH$_3$, CH$_2$} &
                 \parbox[t]{11cm}{C--H stretching} \\
    1740       & \parbox[t]{3cm}{CHO, C=O} &
                 \parbox[t]{11cm}{C=O stretching} \\
    1670       & \parbox[t]{3cm}{CHO} &
                 \parbox[t]{11cm}{ortho-hydroxy, amino and formyl group, chelate H bond} \\
    1650--1540 & \parbox[t]{3cm}{NH$_2$, NH} &
                 \parbox[t]{11cm}{N--H bending} \\
    1467       & \parbox[t]{3cm}{CH$_3$} &
                 \parbox[t]{11cm}{bending mode, perpendicular to the plane} \\
    1390       & \parbox[t]{3cm}{C--N} &
                 \parbox[t]{11cm}{aniline-like and tertiary C--N stretching} \\
    1190--1050 & \parbox[t]{3cm}{C--N} &
                 \parbox[t]{11cm}{secondary, primary C--N stretching} \\
  \end{tabular}
  \end{ruledtabular}
\end{table*}

\begin{table*}[t]
  \caption{\label{tab:FTIR-ND30}FTIR data measured on the ND30 FNDs with various surface terminations. The observed vibration bands, the associated functional groups, and the respective vibration modes are given.}
  \begin{ruledtabular}
  \begin{tabular}{l l l}
    band (cm$^{-1}$) & functional groups & vibration modes \\ \colrule
    \multicolumn{3}{l}{ND30 as-received} \\
    3550--3090 & \parbox[t]{3cm}{COOH, OH} &
                 \parbox[t]{11cm}{O--H stretching} \\
    2980--2815 & \parbox[t]{3cm}{CH$_3$, CH$_2$} &
                 \parbox[t]{11cm}{C--H stretching} \\
    1860--1560 & \parbox[t]{3cm}{CO, CHO, OH} &
                 \parbox[t]{11cm}{C=O from carboxyl group (monomer, dimer), acid anhydride (symmetric and asymmetric), O--H in intramolecular H-bond, chelate bond} \\
    1471       & \parbox[t]{3cm}{CH$_3$, C--C} &
                 \parbox[t]{11cm}{bending mode, perpendicular to the plane, C--C skeletal} \\
    1300--1000 & \parbox[t]{3cm}{SO$_3$X, C--O--C, C--O} &
                 \parbox[t]{11cm}{from sulfonate; symmetric and asymmetric C--O--C stretching; C--O stretching (primary, secondary, tertiary)} \\
    \colrule
    \multicolumn{3}{l}{ND30 $-$OH terminated} \\
    3500--3120 & \parbox[t]{3cm}{OH} &
                 \parbox[t]{11cm}{O--H stretching} \\
    2980--2790 & \parbox[t]{3cm}{CH$_3$, CH$_2$} &
                 \parbox[t]{11cm}{C--H stretching} \\
    2100--1930 & \parbox[t]{3cm}{OH} &
                 \parbox[t]{11cm}{intramolecular H bond, chelate bond} \\
    1820--1700 & \parbox[t]{3cm}{CHO, CO} &
                 \parbox[t]{11cm}{unreacted CHO and CO groups, chelate H bond} \\
    1680--1550 & \parbox[t]{3cm}{CHO, CO} &
                 \parbox[t]{11cm}{C=O stretching; unreacted CHO and O--H in chelate H bond (ortho position), adsorbed water} \\
    1450       & \parbox[t]{3cm}{CH$_3$} &
                 \parbox[t]{11cm}{bending mode, perpendicular to the plane} \\
    1300--1100 & \parbox[t]{3cm}{C--O, C--C} &
                 \parbox[t]{11cm}{C--O stretching modes (primary, secondary, tertiary); skeletal vibration} \\
    710        & \parbox[t]{3cm}{OH} &
                 \parbox[t]{11cm}{O--H deformation mode in H-bond} \\
    \colrule
    \multicolumn{3}{l}{ND30 $-$NH$_2$ terminated} \\
    3550--3300 & \parbox[t]{3cm}{NH$_2$, NH} &
                 \parbox[t]{11cm}{N--H stretching} \\
    2980--2825 & \parbox[t]{3cm}{CH$_3$, CH$_2$} &
                 \parbox[t]{11cm}{C--H stretching} \\
    1733       & \parbox[t]{3cm}{CONH$_2$, CONH} &
                 \parbox[t]{11cm}{unreacted amide groups (cyclic amide, lactam)} \\
    1640--1560 & \parbox[t]{3cm}{NH$_2$, NH} &
                 \parbox[t]{11cm}{N--H bending} \\
    1450       & \parbox[t]{3cm}{CH$_3$} &
                 \parbox[t]{11cm}{bending mode, perpendicular to the plane} \\
    1390--1050 & \parbox[t]{3cm}{C--N} &
                 \parbox[t]{11cm}{aniline-like and tertiary, secondary, primary C--N stretching} \\
    705        & \parbox[t]{3cm}{NH$_2$} &
                 \parbox[t]{11cm}{N--H bending perpendicular to the plane} \\
  \end{tabular}
  \end{ruledtabular}
\end{table*}

\begin{table*}[t]
  \caption{\label{tab:FTIR-ND40}FTIR data measured on the ND40 FNDs with various surface terminations. The observed vibration bands, the associated functional groups, and the respective vibration modes are given.}
  \begin{ruledtabular}
  \begin{tabular}{l l l}
    band (cm$^{-1}$) & functional groups & vibration modes \\ \colrule
    \multicolumn{3}{l}{ND40 as-received} \\
    3600--3010 & \parbox[t]{3cm}{COOH, OH} &
                 \parbox[t]{11cm}{O--H stretching} \\
    2970--2840 & \parbox[t]{3cm}{CH$_3$, CH$_2$} &
                 \parbox[t]{11cm}{C--H stretching} \\
    1860--1620 & \parbox[t]{3cm}{CO, CHO, OH} &
                 \parbox[t]{11cm}{C=O from carboxyl group (monomer, dimer), acid anhydride (symmetric and asymmetric), O--H in intramolecular H-bond, chelate bond} \\
    1473       & \parbox[t]{3cm}{CH$_3$, C--C} &
                 \parbox[t]{11cm}{bending mode, perpendicular to the plane, C--C skeletal} \\
    1350       & \parbox[t]{3cm}{SO$_3$X, C--O--C, C--O} &
                 \parbox[t]{11cm}{from sulfonate; symmetric and asymmetric C--O--C stretching; C--O stretching (primary, secondary, tertiary)} \\
    \colrule
    \multicolumn{3}{l}{ND40 $-$OH terminated} \\
    3520--3100 & \parbox[t]{3cm}{OH} &
                 \parbox[t]{11cm}{O--H stretching} \\
    2980--2790 & \parbox[t]{3cm}{CH$_3$, CH$_2$} &
                 \parbox[t]{11cm}{C--H stretching} \\
    1720--1600 & \parbox[t]{3cm}{CHO, CO} &
                 \parbox[t]{11cm}{unreacted CHO and CO groups; chelate H bond} \\
    1680--1550 & \parbox[t]{3cm}{CHO, CO, C=C, OH} &
                 \parbox[t]{11cm}{C=O stretching; unreacted CHO and O--H in chelate H bond (ortho position); adsorbed water; C=C stretching} \\
    1450       & \parbox[t]{3cm}{CH$_3$} &
                 \parbox[t]{11cm}{bending mode, perpendicular to the plane} \\
    1360       & \parbox[t]{3cm}{OH} &
                 \parbox[t]{11cm}{O--H bending} \\
    1290--1220 & \parbox[t]{3cm}{C--O--C, C--O} &
                 \parbox[t]{11cm}{C--O stretching (secondary, tertiary)} \\
    1080--1000 & \parbox[t]{3cm}{C--O} &
                 \parbox[t]{11cm}{C--O stretching (primary, secondary)} \\
    720        & \parbox[t]{3cm}{OH} &
                 \parbox[t]{11cm}{O--H deformation mode in H-bond} \\
    \colrule
    \multicolumn{3}{l}{ND40 $-$NH$_2$ terminated} \\
    3540--3090 & \parbox[t]{3cm}{NH$_2$, NH} &
                 \parbox[t]{11cm}{N--H stretching} \\
    2980--2830 & \parbox[t]{3cm}{CH$_3$, CH$_2$} &
                 \parbox[t]{11cm}{C--H stretching} \\
    1845--1580 & \parbox[t]{3cm}{CONH$_2$, CONH, NH$_2$, NH} &
                 \parbox[t]{11cm}{unreacted amide groups (cyclic amide, lactam); N--H bending} \\
    1450       & \parbox[t]{3cm}{CH$_3$} &
                 \parbox[t]{11cm}{bending mode, perpendicular to the plane} \\
    1410       & \parbox[t]{3cm}{amide III band} &
                 \parbox[t]{11cm}{C--N stretching, NH$_2$ bending, C=O stretching} \\
    1320--1000 & \parbox[t]{3cm}{C--N} &
                 \parbox[t]{11cm}{aniline-like and tertiary, secondary, primary C--N stretching} \\
    806        & \parbox[t]{3cm}{NH$_2$} &
                 \parbox[t]{11cm}{N--H bending perpendicular to the plane} \\
  \end{tabular}
  \end{ruledtabular}
\end{table*}

\begin{table*}[t]
  \caption{\label{tab:FTIR-ND50}FTIR data measured on the ND50 FNDs with various surface terminations. The observed vibration bands, the associated functional groups, and the respective vibration modes are given.}
  \begin{ruledtabular}
  \begin{tabular}{l l l}
    band (cm$^{-1}$) & functional groups & vibration modes \\ \colrule
    \multicolumn{3}{l}{ND50 as-received} \\
    3450--3000 & \parbox[t]{3cm}{COOH, OH} &
                 \parbox[t]{11cm}{O--H stretching} \\
    2925, 2855 & \parbox[t]{3cm}{CH$_3$, CH$_2$} &
                 \parbox[t]{11cm}{C--H stretching} \\
    1835--1605 & \parbox[t]{3cm}{COOH, CO, CHO, C=C} &
                 \parbox[t]{11cm}{C=O from carboxyl group (monomer, dimer), acid anhydride (symmetric and asymmetric), H bond in chelate bond, C=C stretching} \\
    1522       & \parbox[t]{3cm}{C=O, C=C, C--C} &
                 \parbox[t]{11cm}{skeletal vibrations; C=O from diketone} \\
    1270--1130 & \parbox[t]{3cm}{C--O} &
                 \parbox[t]{11cm}{tertiary, secondary and aryl-like C--O; asym./sym. C--O--C} \\
    1390       & \parbox[t]{3cm}{C--O} &
                 \parbox[t]{11cm}{ stretching (tertiary)} \\
    1190--1050 & \parbox[t]{3cm}{C--O} &
                 \parbox[t]{11cm}{stretching (primary, secondary)} \\
    890--835   & \parbox[t]{3cm}{C--O--C} &
                 \parbox[t]{11cm}{asymmetric/symmetric stretching from dioxane and epoxide-like C--O--C units} \\
    \colrule
    \multicolumn{3}{l}{ND50 $-$OH terminated} \\
    3600--3000 & \parbox[t]{3cm}{OH} &
                 \parbox[t]{11cm}{O--H stretching} \\
    2928, 2855 & \parbox[t]{3cm}{CH$_3$, CH$_2$} &
                 \parbox[t]{11cm}{C--H stretching} \\
    1230--1000 & \parbox[t]{3cm}{C--O} &
                 \parbox[t]{11cm}{tertiary, secondary and aryl-like C--O} \\
    \colrule
    \multicolumn{3}{l}{ND50 $-$NH$_2$ terminated} \\
    3250       & \parbox[t]{3cm}{NH$_2$, NH} &
                 \parbox[t]{11cm}{N--H associated form} \\
    1760--1680 & \parbox[t]{3cm}{CHO, C=O} &
                 \parbox[t]{11cm}{ortho-hydroxy, amino and formyl group; chelate H bond} \\
    1220--1080 & \parbox[t]{3cm}{C--N} &
                 \parbox[t]{11cm}{C--N stretching} \\
    800        & \parbox[t]{3cm}{NH$_2$} &
                 \parbox[t]{11cm}{N--H bending mode, perpendicular to the plane} \\
  \end{tabular}
  \end{ruledtabular}
\end{table*}

\begin{table*}[t]
  \caption{\label{tab:FTIR-ND70}FTIR data measured on the ND70 FNDs with various surface terminations. The observed vibration bands, the associated functional groups, and the respective vibration modes are given.}
  \begin{ruledtabular}
  \begin{tabular}{l l l}
    band (cm$^{-1}$) & functional groups & vibration modes \\ \colrule
    \multicolumn{3}{l}{ND70 as-received} \\
    3530--3090 & \parbox[t]{3cm}{COOH, OH} &
                 \parbox[t]{11cm}{O--H stretching} \\
    2980--2830 & \parbox[t]{3cm}{CH$_3$, CH$_2$} &
                 \parbox[t]{11cm}{C--H stretching} \\
    1980--1800 & \parbox[t]{3cm}{OH, CO, CHO} &
                 \parbox[t]{11cm}{C=O from carboxyl group (monomer, dimer), acid anhydride (symmetric and asymmetric), O--H in intramolecular H-bond} \\
    1800--1600 & \parbox[t]{3cm}{CO, CHO, OH, C=C} &
                 \parbox[t]{11cm}{C=O from carboxyl group (monomer, dimer), acid anhydride (symmetric and asymmetric), O--H in intramolecular H-bond, chelate bond, C=C stretching} \\
    1455       & \parbox[t]{3cm}{CH$_3$} &
                 \parbox[t]{11cm}{bending mode, perpendicular to the plane, C--C skeletal} \\
    1220--1010 & \parbox[t]{3cm}{SO$_3$X, C--O--C, C--O} &
                 \parbox[t]{11cm}{from sulfonate; symmetric and asymmetric C--O--C stretching; C--O stretching (primary, secondary, tertiary)} \\
    \colrule
    \multicolumn{3}{l}{ND70 $-$OH terminated} \\
    3580--3110 & \parbox[t]{3cm}{OH} &
                 \parbox[t]{11cm}{O--H stretching} \\
    2990--2830 & \parbox[t]{3cm}{CH$_3$, CH$_2$} &
                 \parbox[t]{11cm}{C--H stretching} \\
    1980--1800 & \parbox[t]{3cm}{OH, CO, CHO} &
                 \parbox[t]{11cm}{C=O from carboxyl group (monomer, dimer), acid anhydride (symmetric and asymmetric), O--H in intramolecular H-bond} \\
    1730--1550 & \parbox[t]{3cm}{CHO, CO} &
                 \parbox[t]{11cm}{C=O stretching; unreacted CHO, C=O and O--H in chelate H bond (ortho position); adsorbed water; C=C stretching} \\
    1480, 1440 & \parbox[t]{3cm}{OH, CH$_3$} &
                 \parbox[t]{11cm}{bending mode, perpendicular to the plane} \\
    1300       & \parbox[t]{3cm}{OH} &
                 \parbox[t]{11cm}{O--H bending} \\
    1250, 1160, 1100 &
                 \parbox[t]{3cm}{C--O} &
                 \parbox[t]{11cm}{C--O stretching (primary, secondary, tertiary)} \\
    710        & \parbox[t]{3cm}{OH} &
                 \parbox[t]{11cm}{O--H deformation mode in H-bond} \\
    \colrule
    \multicolumn{3}{l}{ND70 $-$NH$_2$ terminated} \\
    3540--3070 & \parbox[t]{3cm}{NH$_2$, NH} &
                 \parbox[t]{11cm}{N--H stretching} \\
    2975--2800 & \parbox[t]{3cm}{CH$_3$, CH$_2$} &
                 \parbox[t]{11cm}{ortho-hydroxy, amino and formyl group, chelate H bond} \\
    1770--1490 & \parbox[t]{3cm}{CONH$_2$, CONH, NH$_2$, NH} &
                 \parbox[t]{11cm}{unreacted amide groups (cyclic amide, lactam), N--H bending} \\
    1460--1300 & \parbox[t]{3cm}{CH$_3$, CH$_2$, C--N} &
                 \parbox[t]{11cm}{C--H bending mode, perpendicular to the plane; C--N stretching (aniline-like)} \\
    1250       & \parbox[t]{3cm}{C--N} &
                 \parbox[t]{11cm}{tertiary, secondary C--N stretching} \\
    1080--1020 & \parbox[t]{3cm}{C--N} &
                 \parbox[t]{11cm}{secondary, primary C--N stretching} \\
    795        & \parbox[t]{3cm}{NH$_2$} &
                 \parbox[t]{11cm}{N--H bending perpendicular to the plane} \\
  \end{tabular}
  \end{ruledtabular}
\end{table*}

\begin{table*}[t]
  \caption{\label{tab:FTIR-ND90}FTIR data measured on the ND90 FNDs with various surface terminations. The observed vibration bands, the associated functional groups, and the respective vibration modes are given.}
  \begin{ruledtabular}
  \begin{tabular}{l l l}
    band (cm$^{-1}$) & functional groups & vibration modes \\ \colrule
    \multicolumn{3}{l}{ND90 as-received} \\
    3380--3050 & \parbox[t]{3cm}{COOH, OH} &
                 \parbox[t]{11cm}{O--H stretching} \\
    1790       & \parbox[t]{3cm}{C=O} &
                 \parbox[t]{11cm}{C=O from acid anhydride, symmetric and antisymmetric} \\
    1690--1600 & \parbox[t]{3cm}{C=O} &
                 \parbox[t]{11cm}{C=O from carboxyl group} \\
    1480       & \parbox[t]{3cm}{CH$_3$, C--C} &
                 \parbox[t]{11cm}{bending mode, perpendicular to the plane, C--C skeletal vibrations} \\
    1310       & \parbox[t]{3cm}{C--O, S=O} &
                 \parbox[t]{11cm}{symmetric C--O stretching (CO--O) and S=O from sulfonate} \\
    1110       & \parbox[t]{3cm}{C--O, C--O--C} &
                 \parbox[t]{11cm}{C--O stretching from C--OH, anhydride and ether-like} \\
    \colrule
    \multicolumn{3}{l}{ND90 $-$OH terminated} \\
    3600--3070 & \parbox[t]{3cm}{OH} &
                 \parbox[t]{11cm}{O--H stretching} \\
    2960--2845 & \parbox[t]{3cm}{CH$_3$, CH$_2$} &
                 \parbox[t]{11cm}{could be the C=O of an unreacted COOH group, or chelate bond seen by the ortho OH--COOH group; $\beta$-diketo carbonyl moiety} \\
    1110       & \parbox[t]{3cm}{C--O} &
                 \parbox[t]{11cm}{C--O stretching mostly from secondary alcoholic group} \\
    \colrule
    \multicolumn{3}{l}{ND90 $-$NH$_2$ terminated} \\
    3370       & \parbox[t]{3cm}{NH$_2$, NH} &
                 \parbox[t]{11cm}{N--H stretching} \\
    2970--2855 & \parbox[t]{3cm}{CH$_3$, CH$_2$} &
                 \parbox[t]{11cm}{C--H stretching} \\
    1785       & \parbox[t]{3cm}{CHO, C=O} &
                 \parbox[t]{11cm}{ortho-hydroxy/amino and formyl group, chelate H bond} \\
    1650--1560 & \parbox[t]{3cm}{NH$_2$, NH} &
                 \parbox[t]{11cm}{primary and secondary N--H bending mode} \\
    1460       & \parbox[t]{3cm}{CH$_3$, C--C} &
                 \parbox[t]{11cm}{ bending mode, perpendicular to the plane; C--C skeletal mode} \\
    1320       & \parbox[t]{3cm}{C--N} &
                 \parbox[t]{11cm}{aromatic-like and secondary C--N stretching} \\
    1212--1100 & \parbox[t]{3cm}{C--N} &
                 \parbox[t]{11cm}{primary and secondary C--N stretching} \\
    808        & \parbox[t]{3cm}{NH$_2$} &
                 \parbox[t]{11cm}{N--H bending mode, perpendicular to the plane} \\
  \end{tabular}
  \end{ruledtabular}
\end{table*}

\begin{table*}[t]
  \caption{\label{tab:FTIR-ND140}FTIR data measured on the ND140 FNDs with various surface terminations. The observed vibration bands, the associated functional groups, and the respective vibration modes are given.}
  \begin{ruledtabular}
  \begin{tabular}{l l l}
    band (cm$^{-1}$) & functional groups & vibration modes \\ \colrule
    \multicolumn{3}{l}{ND140 as-received} \\
    3600--3060 & \parbox[t]{3cm}{COOH, OH} &
                 \parbox[t]{11cm}{O--H stretching} \\
    2980--2830 & \parbox[t]{3cm}{CH$_3$, CH$_2$} &
                 \parbox[t]{11cm}{C--H stretching} \\
    1820--1620 & \parbox[t]{3cm}{COOH, CO, CHO} &
                 \parbox[t]{11cm}{C=O from carboxyl group (monomer, dimer), acid anhydride (symmetric and asymmetric)} \\
    1460       & \parbox[t]{3cm}{CH$_3$, C--C} &
                 \parbox[t]{11cm}{bending mode, perpendicular to the plane, C--C skeletal} \\
    1350--1120 & \parbox[t]{3cm}{SO$_3$X, C--O--C, C--O} &
                 \parbox[t]{11cm}{from sulfonate; symmetric and asymmetric C--O--C stretching; C--O stretching (secondary, tertiary)} \\
    1180--1000 & \parbox[t]{3cm}{C--O} &
                 \parbox[t]{11cm}{C--O stretching modes (primary, secondary, tertiary)} \\
    \colrule
    \multicolumn{3}{l}{ND140 $-$OH terminated} \\
    3550--3100 & \parbox[t]{3cm}{OH} &
                 \parbox[t]{11cm}{O--H stretching} \\
    2980--2830 & \parbox[t]{3cm}{CH$_3$, CH$_2$} &
                 \parbox[t]{11cm}{C--H stretching} \\
    1740--1600 & \parbox[t]{3cm}{CHO, CO} &
                 \parbox[t]{11cm}{C=O from unreacted carboxyl group (monomer, dimer); CHO formyl group from reduced COOH (C=O symmetric and asymmetric stretching)} \\
    1460       & \parbox[t]{3cm}{CH$_3$} &
                 \parbox[t]{11cm}{bending mode, perpendicular to the plane} \\
    1150--980  & \parbox[t]{3cm}{C--O, C--C} &
                 \parbox[t]{11cm}{C--O stretching modes (primary, secondary, tertiary); skeletal vibration} \\
    \colrule
    \multicolumn{3}{l}{ND140 $-$NH$_2$ terminated} \\
    3500--3080 & \parbox[t]{3cm}{NH$_2$, NH} &
                 \parbox[t]{11cm}{N--H stretching} \\
    2960--2790 & \parbox[t]{3cm}{CH$_3$, CH$_2$} &
                 \parbox[t]{11cm}{C--H stretching} \\
    1730       & \parbox[t]{3cm}{C=O} &
                 \parbox[t]{11cm}{C=O stretching from unreacted amide (lactam, cyclic amide)} \\
    1670       & \parbox[t]{3cm}{CHO, NH} &
                 \parbox[t]{11cm}{ortho-hydroxy, amino and formyl group, chelate H bond} \\
    1580--1540 & \parbox[t]{3cm}{NH$_2$, NH} &
                 \parbox[t]{11cm}{N--H bending} \\
    1457       & \parbox[t]{3cm}{CH$_3$} &
                 \parbox[t]{11cm}{bending mode, perpendicular to the plane} \\
    1390--1160 & \parbox[t]{3cm}{C--N} &
                 \parbox[t]{11cm}{aniline-like and tertiary C--N stretching} \\
    1060--1040 & \parbox[t]{3cm}{C--N} &
                 \parbox[t]{11cm}{secondary amino C--N stretching} \\
  \end{tabular}
  \end{ruledtabular}
\end{table*}


\FloatBarrier
\section{XPS supporting data}
\label{app:XPS}

\begin{figure}
    \includegraphics[width=0.5\textwidth]{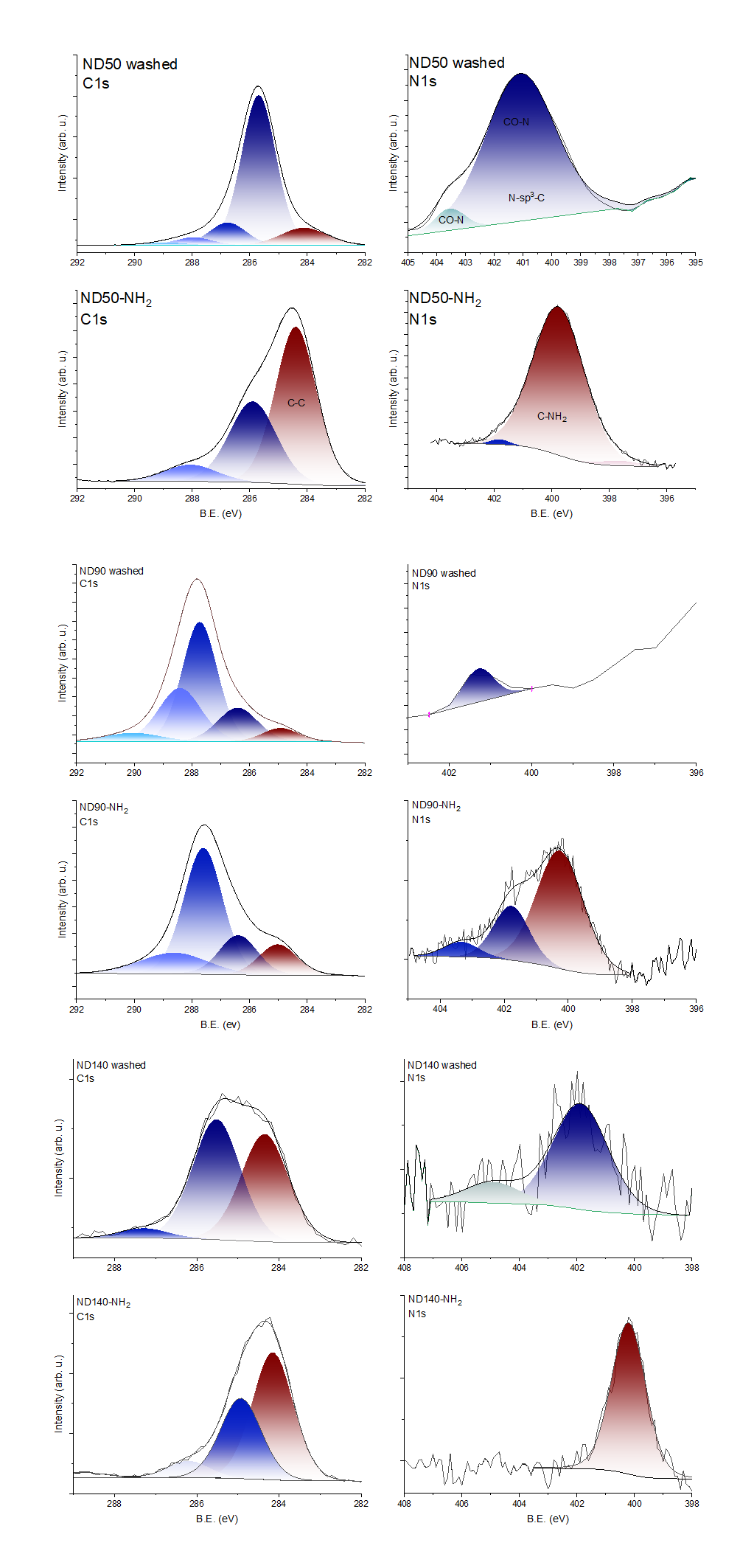}
        \caption{\label{fig:XPS_spectra}The XPS spectra of C and N elements in ND50, ND90 and ND140 FNDs}
    \end{figure}

The XPS spectra of the C and N core levels for the ND50, ND90, and ND140 samples are presented in Fig.~\ref{fig:XPS_spectra}. Please refer to the main text for the detailed interpretation of the peak assignments.

\FloatBarrier
\section{Additional photoluminescence (PL) data}
\label{app:PL}
\begin{figure}
    \includegraphics[width=0.5\textwidth]{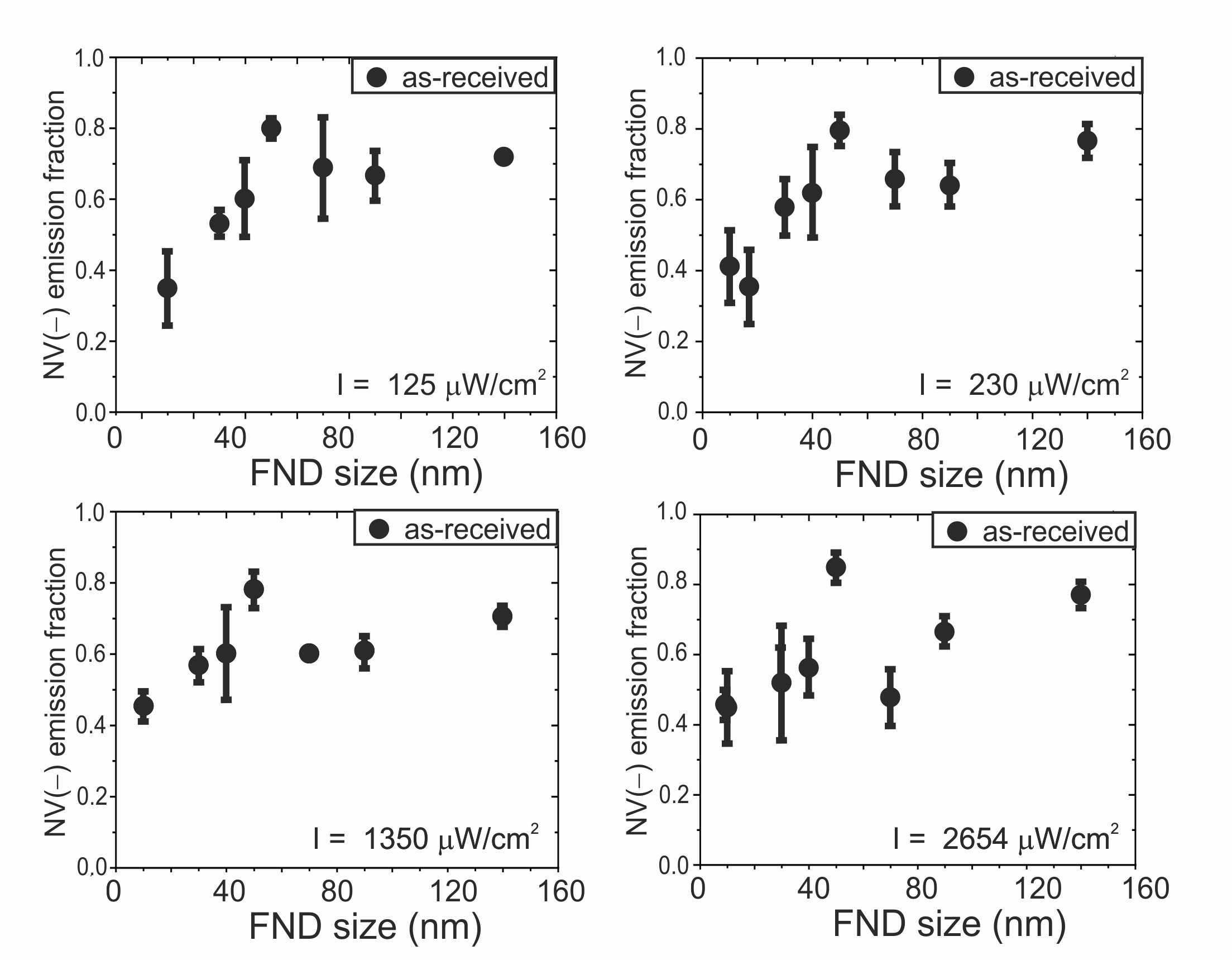}
        \caption{\label{fig:pow-as-received}NV($-$) emission fraction for the as-received FNDs as a function of the applied laser power for different FND sizes.}
    \end{figure}

\begin{figure*}
    \includegraphics[width=0.9\textwidth]{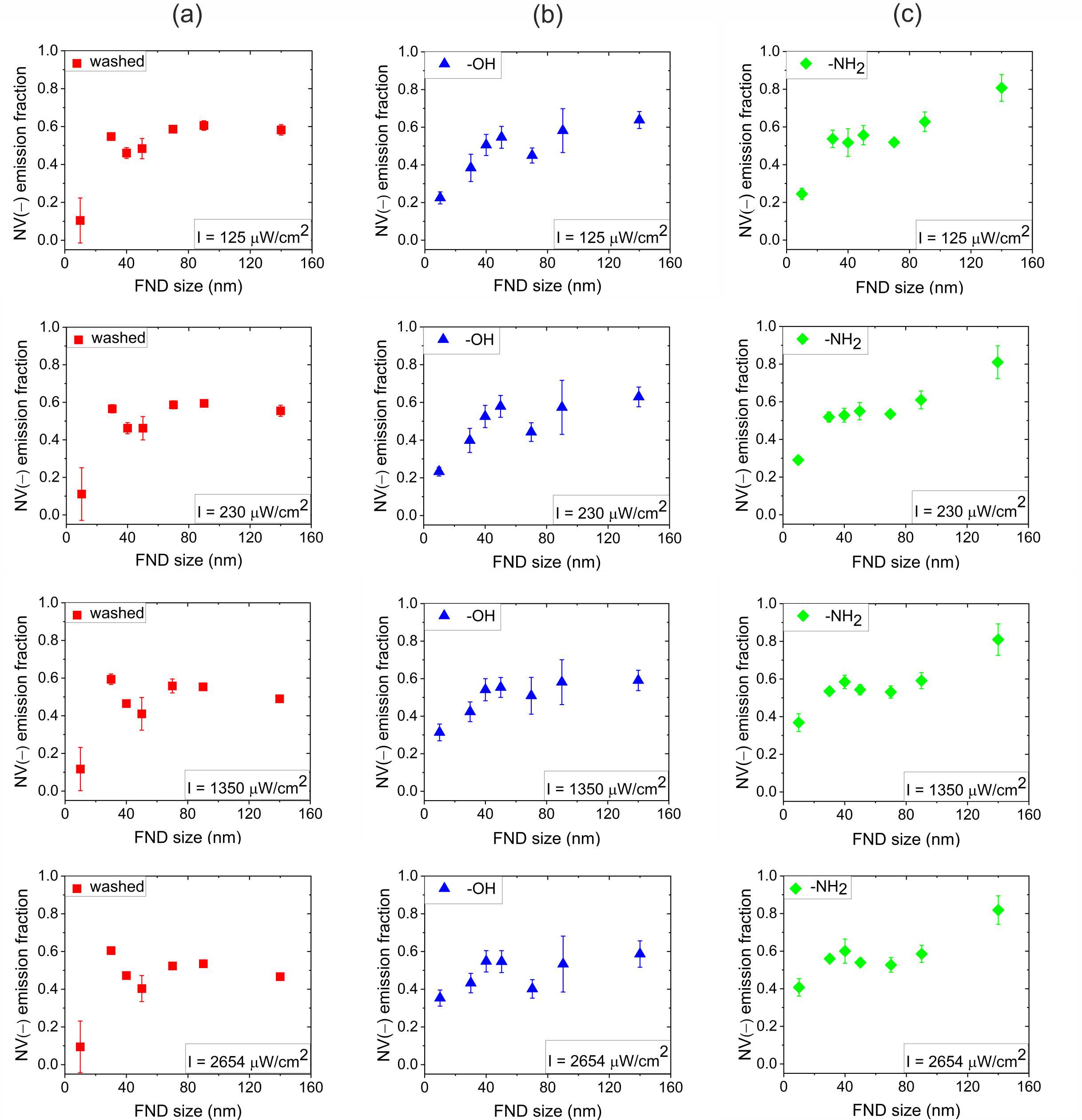}
    \caption{\label{fig:PL_FND_versus_density_size} Photoluminescence spectrum analysis of FNDs. NV($-$) emission fraction for the (a) washed, (b) $-$OH terminated, and (c) $-$NH$_2$ terminated FNDs as a function of the applied laser intensity for different FND sizes.}
    \end{figure*}

The NV($-$) emission fraction was determined for the as-received FNDs as a function of the applied laser power (Fig.~\ref{fig:pow-as-received}). The corresponding results for the washed, $-$OH-terminated, and $-$NH$_2$-terminated FNDs, measured over an extended range of laser powers, are presented in Fig.~\ref{fig:PL_FND_versus_density_size}.

\clearpage

\end{document}